\documentclass[10pt]{article}

\usepackage[
  a4paper,
  top=0.65in, bottom=1.0in, left=0.4in, right=0.4in,
  columnsep=0.41in
]{geometry}

\usepackage{fontspec}
\defaultfontfeatures{Ligatures=TeX}
\newfontfamily\headingfont{Satoshi}[
  Path = fonts/, Extension = .otf,
  UprightFont = *-Medium, ItalicFont = *-MediumItalic
]
\newfontfamily\accentfont{Martian Mono Light 75}[
  Path = fonts/, Extension = .ttf, UprightFont = MartianMono-Light75,
  LetterSpace = 4.0
]

\newlength{\pioarxivshift}
\usepackage{xcolor}
\definecolor{alabaster}{HTML}{F7F3ED}
\definecolor{midnightviolet}{HTML}{423759}
\definecolor{darkmoss}{HTML}{314942}
\definecolor{sunburst}{HTML}{FFA632}
\definecolor{flame}{HTML}{FF8633}
\definecolor{tablerule}{gray}{0.82}

\usepackage{amsmath}
\usepackage{mathastext}
\AtBeginDocument{%
  \DeclareMathSymbol{0}{\mathalpha}{mtletterfont}{`0}%
  \DeclareMathSymbol{1}{\mathalpha}{mtletterfont}{`1}%
  \DeclareMathSymbol{2}{\mathalpha}{mtletterfont}{`2}%
  \DeclareMathSymbol{3}{\mathalpha}{mtletterfont}{`3}%
  \DeclareMathSymbol{4}{\mathalpha}{mtletterfont}{`4}%
  \DeclareMathSymbol{5}{\mathalpha}{mtletterfont}{`5}%
  \DeclareMathSymbol{6}{\mathalpha}{mtletterfont}{`6}%
  \DeclareMathSymbol{7}{\mathalpha}{mtletterfont}{`7}%
  \DeclareMathSymbol{8}{\mathalpha}{mtletterfont}{`8}%
  \DeclareMathSymbol{9}{\mathalpha}{mtletterfont}{`9}%
  \DeclareMathSymbol{\times}{\mathbin}{mtletterfont}{`×}%
}
\usepackage{placeins}
\usepackage{stfloats}
\usepackage{cuted}
\usepackage{needspace}
\usepackage{graphicx}
\usepackage{tikz}
\usepackage{eso-pic}
\usepackage{tcolorbox}
\tcbuselibrary{skins}
\tcbuselibrary{breakable}
\usepackage{titlesec}
\usepackage{ragged2e}
\usepackage{enumitem}
\usepackage{microtype}
\usepackage{booktabs}
\usepackage{tabularx}
\usepackage{colortbl}
\usepackage{multirow}
\usepackage{wrapfig}
\usepackage{multicol}
\usepackage{tocloft}
\usepackage{xurl}
\usepackage[colorlinks=true, linkcolor=midnightviolet, urlcolor=midnightviolet, citecolor=midnightviolet]{hyperref}
\renewcommand{\normalsize}{\fontsize{8}{12}\selectfont}
\AtBeginDocument{\normalsize}
\titleformat{\section}
  {\color{black}\accentfont\fontsize{10}{14}\selectfont}
  {}{0pt}{}
\titlespacing*{\section}{0pt}{1.2em}{0.5em}
\let\piosectionorig\section
\renewcommand{\section}[1]{\piosectionorig{\MakeUppercase{#1}}}

\titleformat{\subsection}
  {\color{black}\headingfont\fontsize{9}{12}\selectfont}
  {}{0pt}{}
\titlespacing*{\subsection}{0pt}{0.9em}{0.3em}

\titleformat{\subsubsection}
  {\color{midnightviolet}\headingfont\fontsize{8.3}{11}\selectfont}
  {}{0pt}{}
\titlespacing*{\subsubsection}{0pt}{0.75em}{0.2em}

\newlength{\piofooterheight}
\newlength{\piomastheadpad}

\AddToShipoutPictureBG{%
  \begin{tikzpicture}[remember picture, overlay, x=1pt, y=1pt]
    \fill[alabaster] (current page.south west) rectangle
      ([yshift=\piofooterheight]current page.south east);
    \node[anchor=south west, inner sep=0pt]
      at ([xshift=0.4in, yshift=0.5\piofooterheight-3pt]current page.south west)
      {\includegraphics[height=6pt]{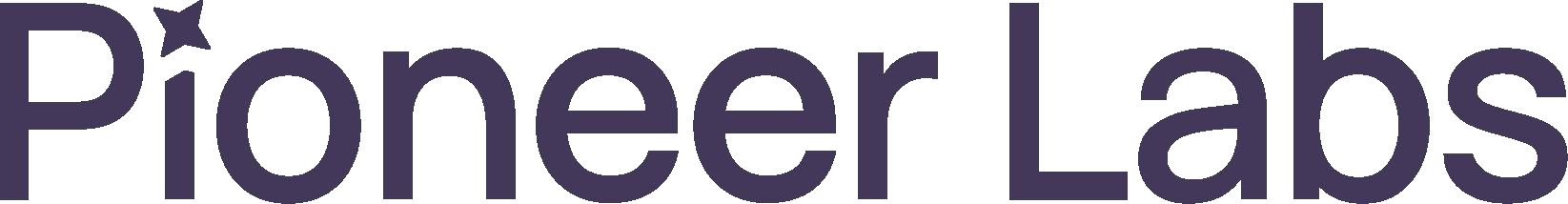}};
    \node[anchor=south east, inner sep=0pt, font=\accentfont\fontsize{5}{6}\selectfont, text=black]
      at ([xshift=-0.4in, yshift=0.5\piofooterheight-2.2pt]current page.south east)
      {\ifnum\value{page}<10 0\fi\thepage};
    \ifnum\value{page}=1
      \coordinate (piombY) at ([yshift=-\piomastheadpad]mastheadbottom);
      \fill[alabaster] (current page.north west) rectangle (piombY -| current page.east);
    \fi
  \end{tikzpicture}%
}

\newtcolorbox{piofigurebox}{
  enhanced,
  colback = white, colframe = midnightviolet,
  boxrule = 0.5pt, arc = 0pt, outer arc = 0pt,
  left = 0pt, right = 0pt, top = 0pt, bottom = 0pt, boxsep = 0pt,
  overlay = {
    \draw[midnightviolet, line width=4pt]
      ([xshift=2pt, yshift=-0.5pt]frame.north west) --
      ([xshift=2pt, yshift=-85pt]frame.north west);
  }
}
\newtcolorbox{piocaptionbox}{
  colback = alabaster, colframe = alabaster, boxrule = 0pt, arc = 0pt,
  left = 9.2pt, right = 10pt, top = 11pt, bottom = 5.6pt,
  width = \linewidth,
}
\newlength{\piofigtopgap}
\newcounter{piofig}
\newcommand{\piofigure}[3]{%
  \begin{figure*}[t]
    \begin{piofigurebox}
      \noindent{\color{midnightviolet}\vrule width 1pt}%
      \begin{minipage}[t]{\dimexpr\linewidth-1pt\relax}
      \par\vspace{\piofigtopgap}
      \hspace{16pt}%
      \begin{minipage}[t]{\dimexpr\linewidth-28pt\relax}
        \centering
        \includegraphics[width=\linewidth, height=0.4\textheight, keepaspectratio]{#1}
      \end{minipage}\par
      \vspace{5pt}
      \begin{piocaptionbox}
        \begingroup\RaggedRight
        {\accentfont\fontsize{7}{9}\selectfont
          \refstepcounter{piofig}Figure~\thepiofig: #2}%
        \addcontentsline{lof}{figure}{\protect\numberline{\thepiofig}#2}%
        \ifx&#3&\else\par\vspace{3pt}{\fontsize{7}{9}\selectfont #3}\fi
        \endgroup
      \end{piocaptionbox}
      \end{minipage}%
    \end{piofigurebox}
  \end{figure*}%
}
\newcommand{\piofigureh}[5][t]{%
  \begin{figure*}[#1]
    \begin{piofigurebox}
      \noindent{\color{midnightviolet}\vrule width 1pt}%
      \begin{minipage}[t]{\dimexpr\linewidth-1pt\relax}
      \par\vspace{\piofigtopgap}
      \hspace{16pt}%
      \begin{minipage}[t]{\dimexpr\linewidth-28pt\relax}
        \centering
        \includegraphics[width=\linewidth, height=#3\textheight, keepaspectratio]{#2}
      \end{minipage}\par
      \vspace{5pt}
      \begin{piocaptionbox}
        \begingroup\RaggedRight
        {\accentfont\fontsize{7}{9}\selectfont
          \refstepcounter{piofig}Figure~\thepiofig: #4}%
        \addcontentsline{lof}{figure}{\protect\numberline{\thepiofig}#4}%
        \ifx&#5&\else\par\vspace{3pt}{\fontsize{7}{9}\selectfont #5}\fi
        \endgroup
      \end{piocaptionbox}
      \end{minipage}%
    \end{piofigurebox}
  \end{figure*}%
}
\newcommand{\piofigurehcol}[5][t]{%
  \begin{figure}[#1]
    \begin{piofigurebox}
      \noindent{\color{midnightviolet}\vrule width 1pt}%
      \begin{minipage}[t]{\dimexpr\linewidth-1pt\relax}
      \par\vspace{\piofigtopgap}
      \hspace{16pt}%
      \begin{minipage}[t]{\dimexpr\linewidth-28pt\relax}
        \centering
        \includegraphics[width=\linewidth, height=#3\textheight, keepaspectratio]{#2}
      \end{minipage}\par
      \vspace{5pt}
      \begin{piocaptionbox}
        \begingroup\RaggedRight
        {\accentfont\fontsize{7}{9}\selectfont
          \refstepcounter{piofig}Figure~\thepiofig: #4}%
        \addcontentsline{lof}{figure}{\protect\numberline{\thepiofig}#4}%
        \ifx&#5&\else\par\vspace{3pt}{\fontsize{7}{9}\selectfont #5}\fi
        \endgroup
      \end{piocaptionbox}
      \end{minipage}%
    \end{piofigurebox}
  \end{figure}%
}

\newcommand{\piofigureinline}[4]{%
  \par\vspace{4pt}%
  \begin{piofigurebox}
    \noindent{\color{midnightviolet}\vrule width 1pt}%
    \begin{minipage}[t]{\dimexpr\linewidth-1pt\relax}
    \par\vspace{\piofigtopgap}
    \hspace{16pt}%
    \begin{minipage}[t]{\dimexpr\linewidth-28pt\relax}
      \centering
      \includegraphics[width=\linewidth, height=#2\textheight, keepaspectratio]{#1}
    \end{minipage}\par
    \vspace{5pt}
    \begin{piocaptionbox}
      \begingroup\RaggedRight
      {\accentfont\fontsize{7}{9}\selectfont
        \refstepcounter{piofig}Figure~\thepiofig: #3}%
      \addcontentsline{lof}{figure}{\protect\numberline{\thepiofig}#3}%
      \ifx&#4&\else\par\vspace{3pt}{\fontsize{7}{9}\selectfont #4}\fi
      \endgroup
    \end{piocaptionbox}
    \end{minipage}%
  \end{piofigurebox}%
}

\newcommand{\piofigurelabeled}[6][t]{%
  \begin{figure*}[#1]
    \begin{piofigurebox}
      \noindent{\color{midnightviolet}\vrule width 1pt}%
      \begin{minipage}[t]{\dimexpr\linewidth-1pt\relax}
      \par\vspace{\piofigtopgap}
      \hspace{16pt}%
      \begin{minipage}[t]{\dimexpr\linewidth-28pt\relax}
        \centering
        \includegraphics[width=\linewidth, height=#4\textheight, keepaspectratio]{#2}
      \end{minipage}\par
      \vspace{5pt}
      \begin{piocaptionbox}
        \begingroup\RaggedRight
        {\accentfont\fontsize{7}{9}\selectfont Figure~#3: #5}%
        \ifx&#6&\else\par\vspace{3pt}{\fontsize{7}{9}\selectfont #6}\fi
        \endgroup
      \end{piocaptionbox}
      \end{minipage}%
    \end{piofigurebox}
  \end{figure*}%
}

\newcounter{piobox}
\newtcolorbox{pioboxenv}[1]{
  enhanced,
  colback     = alabaster,
  colframe    = alabaster,
  boxrule     = 0pt,
  arc         = 0pt, outer arc = 0pt,
  left = 9pt, right = 9pt, top = 8pt, bottom = 8pt,
  borderline west = {2pt}{0pt}{midnightviolet},
  parbox      = false,
  title       = {#1},
  fonttitle   = \accentfont\fontsize{7.5}{10}\selectfont\color{black},
  coltitle    = black,
  colbacktitle= alabaster,
  attach title to upper = {\par\vspace{5pt}},
}

\newenvironment{piobox}[2][t]{%
  \refstepcounter{piobox}%
  \begin{figure*}[#1]%
  \setlength{\parskip}{0.55em}\setlength{\parindent}{0pt}%
  \begin{pioboxenv}{#2}%
}{%
  \end{pioboxenv}%
  \end{figure*}%
}

\newcommand{\piotablecaptionlab}[4][\textwidth]{%
  \begin{tcolorbox}[
      colback=alabaster, colframe=alabaster, boxrule=0pt, arc=0pt,
      left=5.4pt, right=8pt, top=8.4pt, bottom=5.8pt,
      width=#1
    ]
    \RaggedRight\fontsize{8}{11}\selectfont
    {\accentfont\fontsize{8}{11}\selectfont Table~#2. #3}%
    \ifx&#4&\else\par\vspace{3pt}{\fontsize{8}{11}\selectfont #4}\fi
  \end{tcolorbox}%
  \addcontentsline{lot}{table}{\protect\numberline{#2}#3}%
  \vspace{-0.55em}\vspace{-2.6pt}%
}
\newcommand{\piotablecaptionlabx}[4][\textwidth]{%
  \begin{tcolorbox}[
      colback=alabaster, colframe=alabaster, boxrule=0pt, arc=0pt,
      left=5.4pt, right=8pt, top=8.4pt, bottom=5.8pt,
      width=#1
    ]
    \RaggedRight\fontsize{8}{11}\selectfont
    {\accentfont\fontsize{8}{11}\selectfont Table~#2. #3}%
    \ifx&#4&\else\par\vspace{3pt}{\fontsize{8}{11}\selectfont #4}\fi
  \end{tcolorbox}%
  \vspace{-0.55em}\vspace{-2.6pt}%
}
\newcolumntype{L}[1]{>{\RaggedRight\arraybackslash}p{#1}}

\newcommand{\piotablebody}{\fontsize{7}{8.8}\selectfont\setlength{\tabcolsep}{4pt}%
  \renewcommand{\arraystretch}{1.45}}

\newcommand{\piohead}[1]{\cellcolor{midnightviolet}\textcolor{white}{#1}}
\arrayrulecolor{tablerule}

\newcommand{\piodivider}{\noindent{\color{midnightviolet}\rule{0.83\linewidth}{0.25pt}}\par}

\newenvironment{pioneerrefsnum}{%
  \begingroup\fontsize{7.5}{10.5}\selectfont
  \setlength{\parindent}{0pt}
  \begin{enumerate}[leftmargin=1.6em, itemindent=0pt, itemsep=0.35em, label=\arabic*.]
}{%
  \end{enumerate}\endgroup
}

\makeatletter
\newcommand{\piostarttoc}[1]{\@starttoc{#1}}
\makeatother

\newcommand{\piolistheading}[1]{%
  {\accentfont\fontsize{8}{12}\selectfont\color{black}\MakeUppercase{#1}}\par\vspace{4pt}%
}
\newcommand{\pioliststyle}{%
  \RaggedRight\hyphenpenalty=10000\exhyphenpenalty=10000\relax
}

\begin{document}
\twocolumn[
  \hspace*{\pioarxivshift}%
  \begin{minipage}{\dimexpr\linewidth-\pioarxivshift\relax}
  {\raggedright
  \begin{minipage}[t]{0.72\linewidth}
    \raggedright
    {\headingfont\fontsize{22pt}{26pt}\selectfont\color{midnightviolet}
      Mars Seed: A Reference Architecture for Biomanufacturing Habitats from Martian Resources\par}
  \end{minipage}\hfill
  \begin{minipage}[t]{0.24\linewidth}
    \raggedleft\vspace{2pt}
    \includegraphics[width=0.95\linewidth]{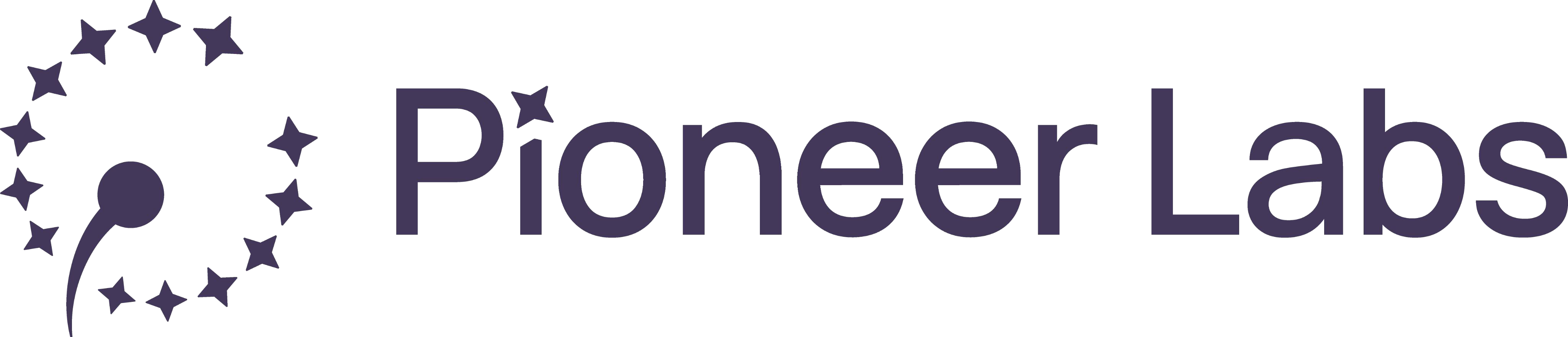}
  \end{minipage}

  \vspace{9pt}
  {\fontsize{10}{13}\selectfont
   \begin{minipage}[t]{0.83\linewidth}\raggedright October 21, 2026, Emeryville, California\end{minipage}\par}
  \vspace{3pt}\piodivider\vspace{3pt}
  {\fontsize{10}{13}\selectfont
   \begin{minipage}[t]{0.83\linewidth}\raggedright \textbf{Key Authors}: Devon A. Stork, Harley Greene, Una Nattermann\end{minipage}\par}
  \vspace{3pt}\piodivider\vspace{3pt}
  {\fontsize{10}{13}\selectfont
   \begin{minipage}[t]{0.83\linewidth}\raggedright \textbf{Pioneer Labs}: Tom Pedersen, Nathan D. Hicks, Max G. Schubert, Fatima R. Martin, Harley Greene, Edward Sukarto, Una Nattermann, Devon A. Stork, Erika A. DeBenedictis\end{minipage}\par}
  \vspace{3pt}\piodivider\vspace{3pt}
  {\fontsize{10}{13}\selectfont
   \begin{minipage}[t]{0.83\linewidth}\raggedright \textbf{Affiliations}: All authors are affiliated with Pioneer Labs, Emeryville, CA 94608, USA\end{minipage}\par}
  \vspace{3pt}\piodivider\vspace{3pt}
  {\fontsize{10}{13}\selectfont
   \begin{minipage}[t]{0.83\linewidth}\raggedright \textbf{Contact}: bioisru@pioneer-labs.org\end{minipage}\par}
  \vspace{3pt}\piodivider
  }%
  {\tikz[remember picture,overlay,x=1pt,y=1pt]\coordinate (mastheadbottom) at (0,0);}%

  \vspace{\piomastheadpad}\vspace{8pt}
  {\accentfont\fontsize{10}{14}\selectfont\color{black} ABSTRACT:}\par\vspace{4pt}
  \fontsize{8}{12}\selectfont

  Human exploration of Mars will require enormous quantities of materials for the first crewed missions and substantially more for permanent habitation, making flexible and scalable technologies for \textit{in situ} resource utilization (ISRU), including biological ISRU, of significant interest. However, no existing mission architecture quantitatively evaluates whether \textit{in situ} biomanufacturing can realistically deliver materials at the scales required for human settlement. Here, we present Mars Seed, an end-to-end reference architecture for using biomanufacturing to produce bioplastics at scale entirely from \textit{in situ} resources on Mars. We provide a complete analysis of a space bioprocess in which feedstocks are generated entirely from Martian resources, converted biologically into polyhydroxybutyrate (PHB) bioplastic, and purified to produce material suitable for constructing human habitats. We show that the upfront power and upmass costs of deploying the equipment required for this bioprocess are recovered within two synodic periods relative to Earth-shipped alternatives. Further, because the process relies entirely on feedstocks sourced from Martian materials, production can continue until equipment failure without requiring additional materials from Earth, enabling scalable conversion of local resources into useful materials. Mars Seed demonstrates that biological ISRU can be mass- and power-competitive with Earth-shipped alternatives, establishing biomanufacturing as a practical component of Mars mission architecture rather than an aspirational capability.

  \vspace{4pt}
  \end{minipage}
]

\begin{strip}
\hspace*{\pioarxivshift}%
\begin{minipage}{\dimexpr\textwidth-\pioarxivshift\relax}
\begin{minipage}[t]{0.575\linewidth}
  \piolistheading{Contents}
  \pioliststyle\piostarttoc{toc}
\end{minipage}\hfill
\begin{minipage}[t]{0.375\linewidth}
  \piolistheading{Figures}
  \pioliststyle\piostarttoc{lof}
  \vspace{9pt}
  \piolistheading{Tables}
  \pioliststyle\piostarttoc{lot}
\end{minipage}
\end{minipage}
\end{strip}
\setcounter{piofig}{0}%
\piofigureh[b]{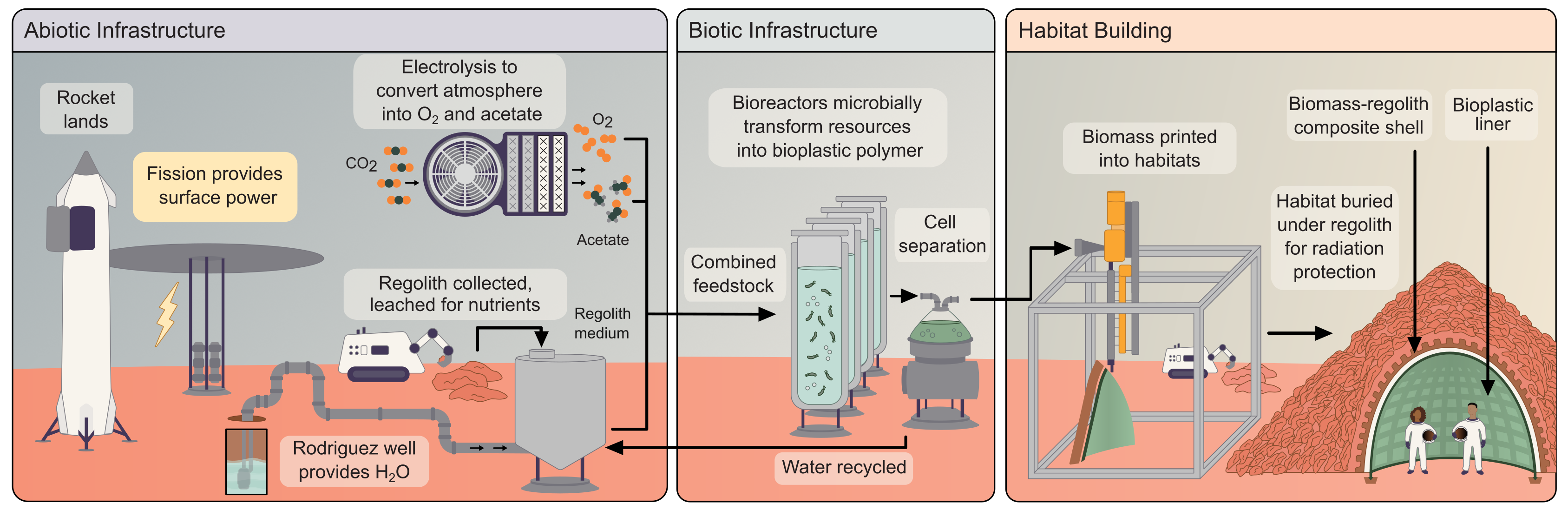}{0.22}{Overview of Mars Seed.}{Artistic representation of Mars Seed biofoundry to produce surface habitats. Abiotic infrastructure creates a feedstock entirely from \textit{in situ} resources. Biotic infrastructure uses a bioprocess to convert this feedstock into bioplastic polymer. Radiation-shielded and pressure-holding surface habitats, safe for human habitation, are built with this bioplastic.}
\null\newpage\null\newpage

\section{Introduction}
An enormous quantity and diversity of materials are required to support the first crewed missions to Mars. In the 1990s, influential reference architectures like Mars Direct\textsuperscript{1} and the Design Reference Mission 1.0\textsuperscript{2} popularized the strategy of predeploying equipment to produce return propellant and other consumables with \textit{in situ} resource utilization (ISRU). These approaches sharply reduce the mass that must be launched from Earth and decrease risk by scheduling the crew’s departure only after vital resources are secure on Mars. Predeployment and ISRU have become staples of modern human Mars mission architectures, with recent designs like the 2021 Strategic Analysis Cycle envisioning launching three equipment payloads over two synodic launch windows prior to human crew departure\textsuperscript{3} and the 2026 NASEM 30-Cargo-300 involving multiple synodic periods of surface activity and cargo delivery prior to a long-stay human mission\textsuperscript{4}. Key ISRU technologies have advanced, most notably MOXIE’s demonstration of oxygen production from Martian atmospheric carbon dioxide\textsuperscript{5}. The need for diverse, scalable \textit{in situ} resource production grows with repeated missions and extended habitation, making expansion of ISRU technology a major NASA priority: currently 18 of NASA’s 63 Moon-to-Mars objectives “involve or enhance ISRU implementation”\textsuperscript{6}. Biological ISRU offers a complementary route to a broad range of products, including structural polymers, food, and pharmaceuticals, and is widely recognized as an enabling technology for sustained human exploration\textsuperscript{7–14}.

However, biological ISRU remains substantially less mature relative to abiotic approaches. A proximal barrier to progress is that there are few concrete proposals that treat biotechnology as a fundamental component. An example of the type of work that is needed is analysis of production of 2,3-butanediol as a potential ascent propellant using a bioprocess on Mars, including quantitative comparison of the mass and power requirements with chemical ISRU alternatives\textsuperscript{15}, although this study analyzes production of a fuel no rocket currently uses and the production still requires Earth-shipped nutrients. Producing a material on Mars does not by itself eliminate dependence on Earth, and end-to-end analysis is necessary to determine whether or not a bioprocess outperforms abiotic alternatives. Terrestrial bioprocesses often involve long supply chains and industrial agriculture to generate feedstocks, which makes translation of terrestrial processes to space difficult. Several proposals exist for using microbes to produce materials in a space context, but require either some Earth supplied nutrients or crew waste for critical nutrients\textsuperscript{8,16,17}. Downstream purification of bioproducts often involves consumables like resins that reduce upmass savings\textsuperscript{18}, and further tie processes to Earth-derived consumables. Replacing these imports with local production carries its own equipment and power costs. An end-to-end reference architecture is needed that accounts for every step from Martian resources to purified product and can establish a framework to calculate when deploying the complete system becomes preferable to shipping the product itself.

Here we describe Mars Seed: the first end-to-end reference architecture centered around biological ISRU as a core technology to produce biobased products. We specify the complete equipment manifest and power requirements of a biofoundry that converts Martian atmosphere, regolith, and ice into purified bioplastic for constructing habitats before crew arrival. 20,000 kg of infrastructure enables construction of 870 m² of habitable space in two synodic periods, more than can be constructed by an equivalent mass of processed materials. Production can subsequently continue without resupply, only limited by equipment lifetime. The Mars Seed biofoundry establishes a scalable habitat construction process, allowing prioritization of scientific payloads and other consumables in subsequent launches.

\newpage

\section{Mars Seed}

\subsection{Overview}
In Mars Seed, an engineered microbe consumes feedstock derived from water, processed atmosphere, and regolith nutrients, and produces bioplastic, which is combined with regolith and turned into habitats in preparation for the arrival of a human crew (\textbf{Figure 1}).

Briefly, a spacecraft containing 20 metric tons (tonnes) of equipment lands at a mid-latitude site on Mars near shallow ice deposits. Robotics unload equipment and set up a 40 kWe nuclear fission reactor, which provides power to assemble the rest of the facilities. This equipment processes \textit{in situ }resources and produces a feedstock composed of water from ice, atmospheric CO\textsubscript{2} processed into acetate, and nitrogen, phosphorus, sulfur, and trace elements from regolith. Four 1,000 L temperature-controlled, pressurized bioreactors are filled with this feedstock and inoculated with an engineered microbe that converts the feedstock into a bioplastic polymer. The bioplastic-containing microbe is separated from the remaining liquid medium and split into two downstream processes. In the first, the bacterial biomass is mixed with regolith to create a composite that is printed to form a shell which is buried under two meters of regolith for radiation and thermal protection. In the second, the polymer is purified from the microbe and is formed into a pressure-containing liner that sits underneath the exterior shell. After about 50 Earth months (1,500 sols) of operation this process produces 870 m\textsuperscript{2} of livable area. At this time, a crew lands nearby with a payload prioritizing exploration and science objectives, and can use the available habitats. The biofoundry continues to produce useful structural materials before, during, and after the crewed visit, and is limited only by equipment lifetime.

We designed this reference architecture to exemplify how biomanufacturing can enable the first crewed missions by offsetting the requisite upmass with biological ISRU to create human-safe habitats. Our rationale going into the design process is as follows:

\begin{enumerate}[leftmargin=1.4em, itemsep=0.25em, topsep=0.25em, label=\arabic*.]
  \item While many biological processes have been described as helpful or necessary for long-term space exploration, we picked biopolymer production for habitats as an example because of its ability to be both useful and cost-effective for crewed habitation and exploration of Mars.
  \item We used the in-development 40 kWe fission surface power concept as a central design point for both expected shipping mass and power output.
  \item We maximized the potential of that power across abiotic, biotic, and habitat equipment; we balanced estimated shipping masses for a 20 tonne total payload.
  \item We referenced the literature and industry standards and performed first-order math to calculate estimates for equipment mass, required power, baseline production, and expected optimizations.
  \item From these calculations, we solved for the rate at which the biopolymer-regolith habitats could be built relative to shipping building materials from Earth.
\end{enumerate}
Below, we describe the equipment and processes in more detail, broken down into three categories: abiotic infrastructure, biotic infrastructure, and habitat building.

\newpage

\subsection{Abiotic Infrastructure}

\subsubsection{Landing, Robotics, and the First 250 Sols}
The first step is choosing a landing site and setting up the biofoundry prior to continuous operation. We assume the landing site is fairly flat, with good access to shallow subsurface ice and regolith. This likely requires a mid-latitude, \textasciitilde{}50° location. After landing, the hardware is unloaded and assembled, either by a precursor human mission, as described in the NASEM campaigns, or with sufficiently capable robotics to assemble and plumb the various subsystems. We aim to rely on autonomous robotics for reliable and safe assembly of equipment, as will any large-scale mission to Mars. For setup of the Mars Seed equipment, we expect full deployment to take approximately 90 sols, with fission surface power assembled in the first 30-40 sols\textsuperscript{19} and assuming a similar amount of time for the initial Rodriguez well melt pool formation and bioreactor inoculation (\textbf{Figure 2}). After the 90 sols, all four bioreactors reach steady-state operation and the first bioplastic is produced. Continuous operation then commences, with the first complete habitat about 210 sols after landing.

\setcounter{piofig}{1}%
\piofigureh[t]{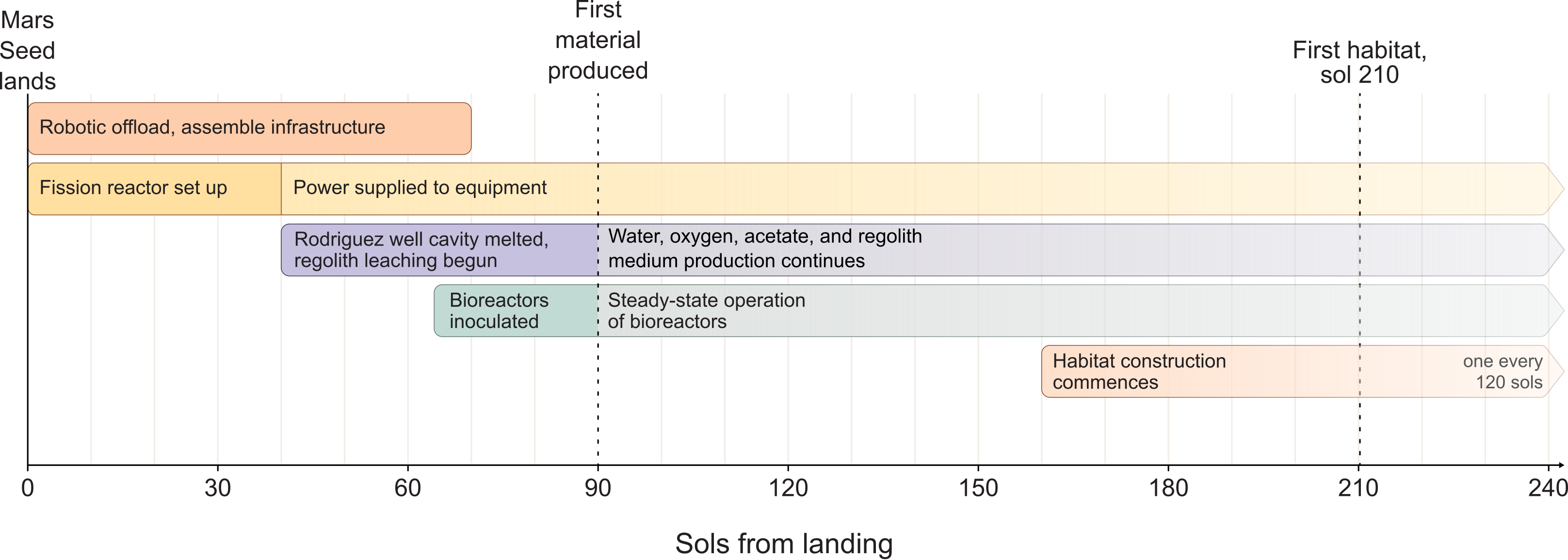}{0.24}{Mars Seed Biofoundry Installation Timeline.}{Timeline for establishing steady-state operations after landing. Fission surface power is established after a 30-40 sol assembly period. The remaining infrastructure is set up and microbe growth commences.}
Beyond setting up the biofoundry, robotics are especially required for transport of regolith, a key \textit{in situ} resource. Mars Seed requires a continuous supply of regolith for three purposes: leaching mineral media (800 kg/day), filler for extruded structural material (50 kg/day), and mounded overburden for radiation shielding (3,200 kg/day), totaling approximately 4 tonnes per day. To provide that hauling capacity, we suggest two rovers, which are based on the Astrolab FLEX, a 500 kg vehicle rated to carry up to 1,600 kg of cargo within 3 m\textsuperscript{3} of payload\textsuperscript{20}. Another design basis could be NASA’s RASSOR systems\textsuperscript{21}, capable of excavating nearly 3,000 kg of regolith per day at even better mass efficiency. The power budget requirements for these systems vary, so we budget at least 1.4 kWe of continuous power for mobility and excavating vehicles based on similar mobility equivalents\textsuperscript{21}. As these robotics also cover other purposes, such as the initial unloading and movement of equipment, we use the mass and power for these haulers as the baseline for autonomous robotics for Mars Seed.

\subsubsection{Surface Power and Water Generation}
Beyond \textit{in situ} regolith, power and water are the basic required inputs for all of the other downstream processes. For surface power, a long-standing NASA design objective is a deployable Fission Surface Power (FSP) system directly applicable to Moon and Mars, with engineering specifications of at least 40 kWe output and mass not exceeding 6,000 kg\textsuperscript{22}. This FSP provides all the electrical power to operate Mars Seed. Such an energy source has also been described in many different surface missions, making it a reliable option as a power source precedent for Mars Seed.

Water on Mars exists in the form of large subsurface ice deposits which must be converted into liquid water. For water generation, we propose a Rodriguez well (Rodwell) approach which drills into these ice deposits, which are common around 40-50° latitude. Subsurface ice deposits on Mars in this area consist of nearly pure ice under a few meters of topcover and at least 100 meters thick\textsuperscript{23,24}, and potentially up to 15 kilometers thick\textsuperscript{25}. Rodwells work by injecting hot water into the well to melt a cavity, and then pumping melted water back up. This technique has decades of field heritage in Greenland and Antarctica, and NASA has found it a promising technique for accessing water on Mars\textsuperscript{26}. The FSP outputs about 130 kW of waste heat\textsuperscript{27}, and we propose using this waste heat to provide the heat for the melt cavity, reducing the electrical power consumption to just that required to maintain pump circulation inside the cavity and transport the meltwater. For a proposed maximum production rate of 400 L of water per day, NASA estimates 11 kW of heat for melting and approximately 0.1 kW of electrical power for the pump\textsuperscript{28}. We allot 750 kg for the Rodwell equipment based on NASA estimates for the mass of a water drill without the heater\textsuperscript{29}, as a heater is not needed if fission waste heat is used. For a size comparison, ten years of 400 L/day water extraction will excavate a melt cavity \textasciitilde{}1,600 m\textsuperscript{3} in size, which is roughly one-tenth the maximum cavity size for the South Polar Water Well in operation at Amundsen-Scott South Pole Station\textsuperscript{30}, indicating that the Rodwell on Mars is of an achievable size.

The biofoundry uses thousands of liters of water daily, however, much of the water can be recycled. About 20 L/day is consumed by electrolysis for acetate production and 2,000 L/day is used to leach regolith and create mineral media for the bioprocess. Once fermentation is complete, the biomass is removed and the spent media can be recycled. For the water recycling, we propose mechanical vapor recompression, where the spent medium is boiled under low pressure, then slightly compressed for recondensation, leaving behind concentrated brine. An 80-90\% recovery rate (i.e. 1,800 L per day) is reasonable given the dilute nature of the spent media, which is estimated to be around 2\% solids based on simulant leaching studies to determine soluble ions\textsuperscript{31,32}. This process is used aboard the ISS to recover water from concentrated urine\textsuperscript{33}, and is a mature technological process in terrestrial applications\textsuperscript{34}, where it has been demonstrated to require at least 16 kWh per thousand liters, or 1.3 kWe of continuous power for the proposed Mars Seed throughput\textsuperscript{35}. We estimate the mass for this water purification system to be around 650 kg, assuming a 1.2 m\textsuperscript{3} container which holds the 6 m\textsuperscript{2} of surface area necessary for efficient evaporation of 75 L/hour of water. The waste stream produced in parallel, composed of 200-400 L/day of post-purification concentrated brine at roughly 20\% solids by weight, might find other uses, potentially as an input stream for electrolysis and production of acid and alkali solutions. With about 1,800 L of water being purified and recycled for leaching, the Rodwell only needs to provide about 200 L of fresh water per day. Including a water recycling system helps satisfy the water requirement for Mars Seed without requiring a massive water ISRU system.

\subsubsection{O\textsubscript{2} and Acetate Production from Atmospheric CO\textsubscript{2}}
The Mars Seed bioprocess is driven by a heterotrophic aerobic microbe, requiring both respirable oxygen and aqueous fixed carbon as energy sources. Both of these are acquired by processing Mars’ thin yet carbon-rich atmosphere, composed of 95\% carbon dioxide (CO\textsubscript{2}). Our proposed carbon source for the bioprocess is acetic acid (C\textsubscript{2}H\textsubscript{4}O\textsubscript{2}), a soluble, 2-carbon feedstock with broad metabolic compatibility that can be manufactured from CO\textsubscript{2} through a two-stage electrolysis process, which also produces the necessary oxygen (O\textsubscript{2}) as a byproduct while requiring only small quantities of water as a secondary input. Other soluble 1- or 2-carbon molecules can be similarly electrochemically derived from CO\textsubscript{2}, such as formate, methanol, and ethanol, but are generally less bio-compatible with common heterotrophic biomanufacturing microbes\textsuperscript{36}. This electrolysis process is already well developed technology for both oxygen and fixed carbon generation. Split across two electrolysis process stages, and sized to optimize biomass production given power constraints, the total daily process and rate of CO\textsubscript{2} conversion is

\begin{center}48.4 kg CO\textsubscript{2} + 19.8 kg H\textsubscript{2}O → 33 kg C\textsubscript{2}H\textsubscript{4}O\textsubscript{2} + 35.2 kg O\textsubscript{2}\end{center}

The first stage directly consumes the CO\textsubscript{2}, splitting it into carbon monoxide (CO) and O\textsubscript{2}, through a solid-oxide electrolysis (SOXE) process:

\begin{center}2 CO\textsubscript{2} → 2 CO + O\textsubscript{2}\end{center}

This generates oxygen for the aerobic bioprocess, much as it would provide oxygen to fill a surface habitat for humans. Critically, this process has been demonstrated aboard the Perseverance rover, called MOXIE, providing an established technology readiness level (TRL) 9 oxygen ISRU system\textsuperscript{5,37} as precedent. Based on previously documented scaling estimates for MOXIE\textsuperscript{38}, we allot the first electrolysis stage a continuous power draw of 6.6 kW and a mass of 850 kg, including the power and mass necessary to pump down the CO\textsubscript{2} from the atmosphere.

In the original MOXIE design, the CO generated is treated as a waste stream and vented out of the system. Here, in a second stage electrolysis, the CO waste product is mixed with water to produce acetate and oxygen:

\begin{center}2 CO + 2 H\textsubscript{2}O →  C\textsubscript{2}H\textsubscript{4}O\textsubscript{2} + O\textsubscript{2}\end{center}

This is based on a commercial industrial process that reduces CO to acetic acid inside an aqueous alkaline electrolyzer. CO reduction to acetate has been demonstrated at about 13 kWh/kg at the process scale\textsuperscript{39}, or 17 kW of continuous power for our proposed throughput. Electrolysis systems to accomplish acetate production at commercial scale have been designed for use in electro-agriculture\textsuperscript{40,41}, with plans for a mobile 100-kW unit\textsuperscript{42}. However, the aqueous alkaline electrolyzer style requires 2 M potassium hydroxide (KOH), with the acetic acid produced as a potassium acetate salt, which must be converted back to acetic acid. This has been demonstrated with a bipolar membrane to regenerate the KOH anolyte solution at an energy cost of 1.9 kWh per kg of acetic acid, or 2.5 kW of continuous power\textsuperscript{43}. Alternatively, a similar CO electrolysis setup can be coupled to an acid-base separation layer for continuous generation of pure acetic acid from carbon monoxide\textsuperscript{44}, though the technology is not as well developed. The total continuous energy draw of stage two would be approximately 20 kW based on existing demonstrations. With a requirement for the electrolyzer stack, radiators, pumps, electrolyte, and bipolar membrane system, we allot 2,750 kg for the full process based on scaling up of existing equipment. Furthermore, the electrolyzer and membrane stacks have the potential to break down over extended cycles, so we allocate an additional 2,500 kg for spares and plumbing, and the system should be designed so the associated robotics can facilitate repairs autonomously. Acetate production from CO remains an area of research, which we expect to only improve with continued investment and technology development.

Taken together, both electrolysis stages one and two are sized to generate the required amount of acetate to feed the four 1,000 L bioreactors each day. At this production capability, the 35 kg O\textsubscript{2} generated in both processes is in excess, as only 11 kg O\textsubscript{2} per day is needed for the bioprocess. The excess O\textsubscript{2} can be used for many purposes, including filling the formed habitats. While costly in mass and power, this system advantageously uses the CO generated from precedent \textit{in situ} O\textsubscript{2} production systems that would typically be an unused waste stream. Similarly, acetate poses as a versatile feedstock for future biological ISRU missions.

\subsubsection{Inorganic Nutrients and Medium from Leached Regolith}
Beyond water, oxygen, and carbon, life requires inorganic nutrients such as nitrogen and phosphorus, and these must be supplied to the bioreactor in sufficient quantities. Decades of Mars rover and lander experiments have demonstrated that these elements are present in the Martian regolith. For Mars Seed, we propose mixing Martian regolith with water in set ratios to leach the necessary inorganic nutrients into solution. In previous work, we developed a defined medium that accurately simulates the soluble chemistry of leached regolith, based on terrestrial leachate data and Mars rover data, and show that mixing 400 grams of regolith with a liter of water and separating the insoluble material will produce a mineral medium capable of supporting high-titer growth of several bioproduction organisms\textsuperscript{32}. Critically, this medium will contain perchlorate, a known soluble stressor in martian regolith, at about 30 mM\textsuperscript{45}. While impactful, many organisms can tolerate that amount of perchlorate\textsuperscript{46}, including the engineered organism designed for Mars Seed\textsuperscript{47}. Producing 2,000 L of culture media per day requires 800 kg of regolith per day. The regolith is added to a simple tank, where it is mixed with water and allowed to settle, after which the aqueous phase is pumped out. Assuming two 1,000 L batches per day, a simple 1,200 L tank with an impeller system and a release hatch to expel spent regolith would be sufficient (\textbf{Figure 3}). Such a system may require filtration to remove micrometer-scale regolith particles from the media; a solid ceramic filter with occasional backwashing is one reliable and reusable engineering precaution that could be included\textsuperscript{48}. Mars Seed turns abundant \textit{in situ} regolith into a scalable culture medium, an aqueous environment containing all the necessary inorganics for the bioprocess.

The mass and power requirement for similar setups on Earth, such as leaching tanks in mineral processing operations\textsuperscript{49} or even bioleaching\textsuperscript{50}, are dominated by the milling requirement\textsuperscript{51}, but the near-surface globally distributed regolith fines available may not require further size reduction\textsuperscript{52}. Therefore, the primary power requirement is stirring the 1,200 L tank with 400 kg of regolith to maintain complete suspension, which should require approximately 0.5 kW under Martian gravity\textsuperscript{53}. A simple system like a leaching tank is sufficient to harvest the necessary nutrients from the regolith, as opposed to complex chemical processes that would further convert the regolith and likely cost more mass and power than allotted here.

\setcounter{piofig}{2}%
\piofigureinline{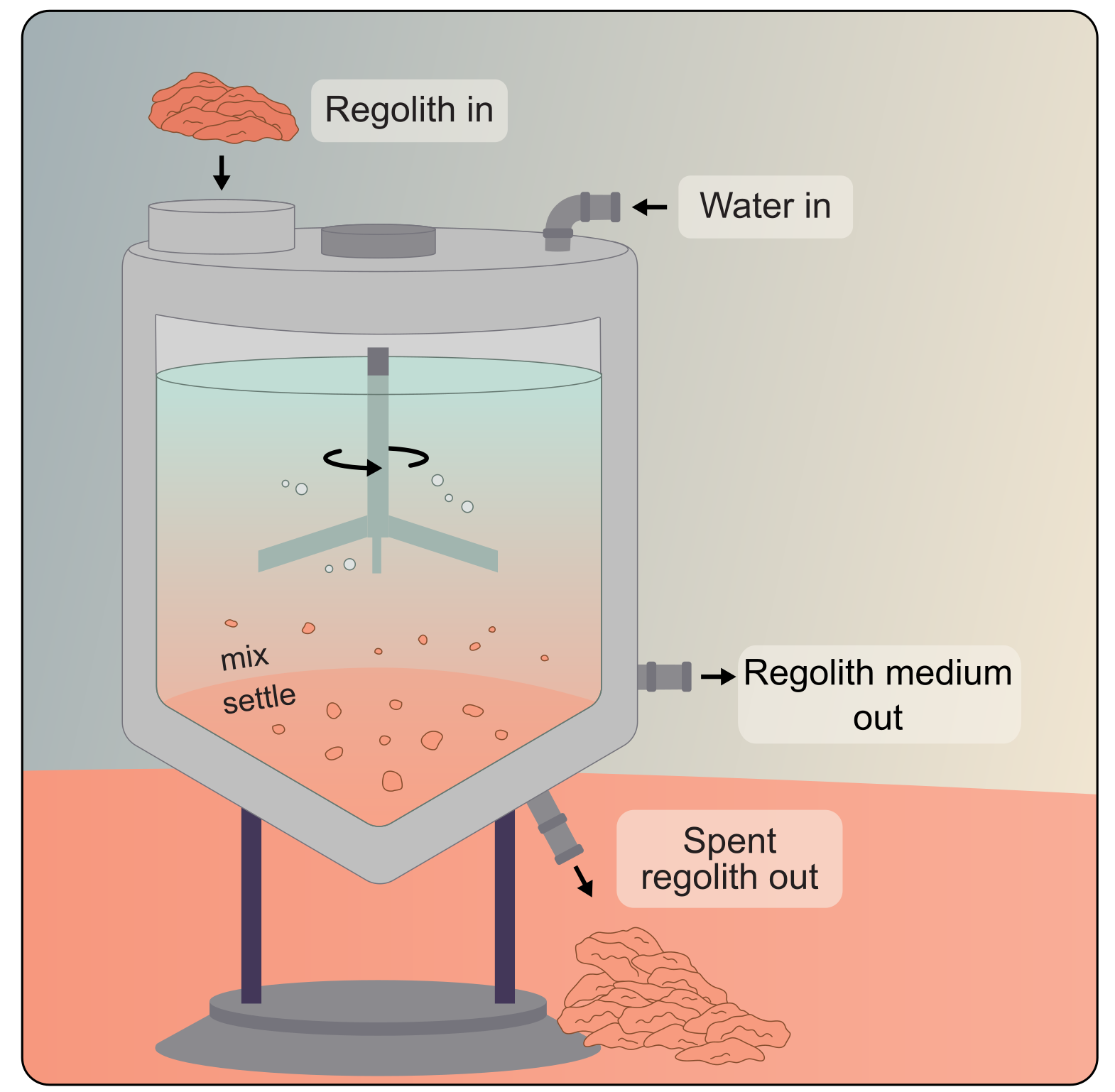}{0.3}{Regolith Leaching Tank Design.}{Artistic representation of a proposed regolith leaching tank. The 1,200 L tank is filled with 1,000 L of water and 400 kg of regolith. The suspension is mixed with an impeller. The aqueous portion is pumped into a bioreactor and insoluble particles are discharged. The process repeats twice a day.}

\subsection{Biotic Infrastructure}
\begin{piobox}[t]{Planetary Protection}
\looseness=-1
One of the aims of planetary protection is to protect extant life on Mars, if it exists, from forward contamination of invasive species from Earth. The NASA policies that governed the era of rovers take the approach of full sterilization of all spacecraft and minimizing the total spore count, and, taken literally, would bar not only this proposed payload in Mars Seed but also any Mars-bound crew, since no astronaut is microbially sterile. Efforts are underway by the Committee on Space Research (COSPAR)’s Planetary Protection Committee and others to transition to new frameworks that assist in decision making during the upcoming era of human spaceflight on Mars. Unlike the human microbiome, which carries unknown microbes and thus unknown risks, clonal organisms used in biomanufacturing can have their risk of becoming invasive quantified exactly. Previously, we have presented a first example of doing so, and demonstrated that the microbes used in Mars Seed, which require a mixed, temperature-controlled, pressurized bioreactor to grow, would be rapidly sterilized if exposed on Mars, and have a propagation probability to \textasciitilde{}10\textsuperscript{-24} per cell, well inside the proposed compliance 10\textsuperscript{-4} risk threshold before considering any equipment-based and containment safeguards\textsuperscript{54}. Well-studied biomanufacturing organisms present a planetary protection risk that is more tractable to characterize relative to the diverse, incompletely characterized microbiome of a human crew.
\end{piobox}

\subsubsection{Microbial Process to Produce Bioplastic Polymers}
Mars Seed uses microbial fermentation to convert Martian resources into bioplastic polymers for use in habitat construction without reliance on Earth-supplied nutrients. The bioprocess used in Mars Seed is well studied: microbes like \textit{Cupriavidus necator} can accumulate bioplastic polymers such as polyhydroxyalkanoates (PHAs) at up to \textasciitilde{}90\% dry cell mass in as little as 24 to 48 hours, depending on the strain and medium conditions, as a form of biological energy storage\textsuperscript{55}. On Earth, microbially-derived bioplastic is generated at industrial scales for use in consumer goods, using nutrients like sugars or agricultural waste streams, as well as bulk inorganic nutrients like ammonium\textsuperscript{55–57}. In contrast, Mars Seed is designed as a fully \textit{in situ} biological process that does not rely on crew presence, Earth-shipped consumables, or Earth-derived nutrients while still generating bioplastic at a meaningful scale.

We have previously shown that the engineered organism sPL.001 (an evolved \textit{C. necator} strain) is capable of aerobic growth and bioplastic production under Martian conditions, using leached Martian regolith as a source of inorganic nutrients and acetate as a carbon and energy source\textsuperscript{47}. This bioprocess must be carried out in a mixed, oxygenated, and warmed system to ensure rapid growth. Optimized fed-batch fermentations, where nutrients are continually supplied, with the same base microorganism and carbon source from Mars Seed produce about 5 g/L polyhydroxybutyrate (PHB), a type of PHA polymer, from about 16.7 g/L of acetate over a 29-hour run at a 5 L scale\textsuperscript{58}. The final cell weight is about 70\% PHB, meaning the final culture has a density of 7 grams of biomass per liter. Assuming similar productivity levels, with 2,000 L/day of mineral media and 33 kg/day of acetic acid, we expect to produce about 14 kg/day of biomass containing 10 kg of PHB for downstream processes, an efficiency of 0.3 grams PHB per gram of input acetate. This process also requires approximately 11 kg/day of O\textsubscript{2} to respire the carbon, provided by the electrolysis process. After the growth process, the microbe will have depleted all the input nutrients, requiring fresh batches of acetate, oxygen, and regolith medium for the subsequent cycles. The fresh media is inoculated with sPL.001 and the bioplastic production continues.

Microbial fermentation is typically housed in a growth container, called a bioreactor. Standard bioreactor designs\textsuperscript{59,60} would only need minor modifications or redesigns to be compatible with Mars-surface functionality. For hardware, we propose a simple bioreactor with minimal moving parts and maintenance requirements, with a per-unit mass and power draw of 550 kg and 0.4 kW (\textbf{Figure 4}), meaning 2,200 kg and 1.6 kW for the four-unit system. The mass is based on using a tall and thin design, about 3 meters tall and a 0.75 meter diameter, to hold 1,000 L of media, with mixing and oxygenation accomplished by sparging gas through a central column, much like an airlift bioreactor. Initial calculations suggest that the airlift mechanism would function in Mars’ lower gravity, requiring a throughput of 50 liters/minute of gas. Each bioreactor has a three millimeter outer shell of 316 stainless steel, sufficient to contain the internal pressure of 110 kPa at the bottom of the reactor. This would mass approximately 200 kg, with the remaining 350 kg per bioreactor allocated for pumps, instrumentation, and a carbon fiber skid. Alternative materials, such as titanium or composites, may further decrease the mass requirement. The power drives the oxygen recirculation and liquid pumping. Further features of the bioreactors include pH and temperature control. Here, pH is controlled by metered addition of acetic acid to counteract alkalinization by metabolism of acetic acid and temperature is controlled by circulating cooled water through a jacket. This bioreactor system could be designed without any hardware requirements for self-sterilization methods, like UV or steam sterilization, due to the decreased risk of external contamination on Mars. All together, the four 1,000 L bioreactors draw heavily on terrestrial precedent and highlight how traditional bioprocess equipment can translate to a space context.

\setcounter{piofig}{3}%
\piofigureinline{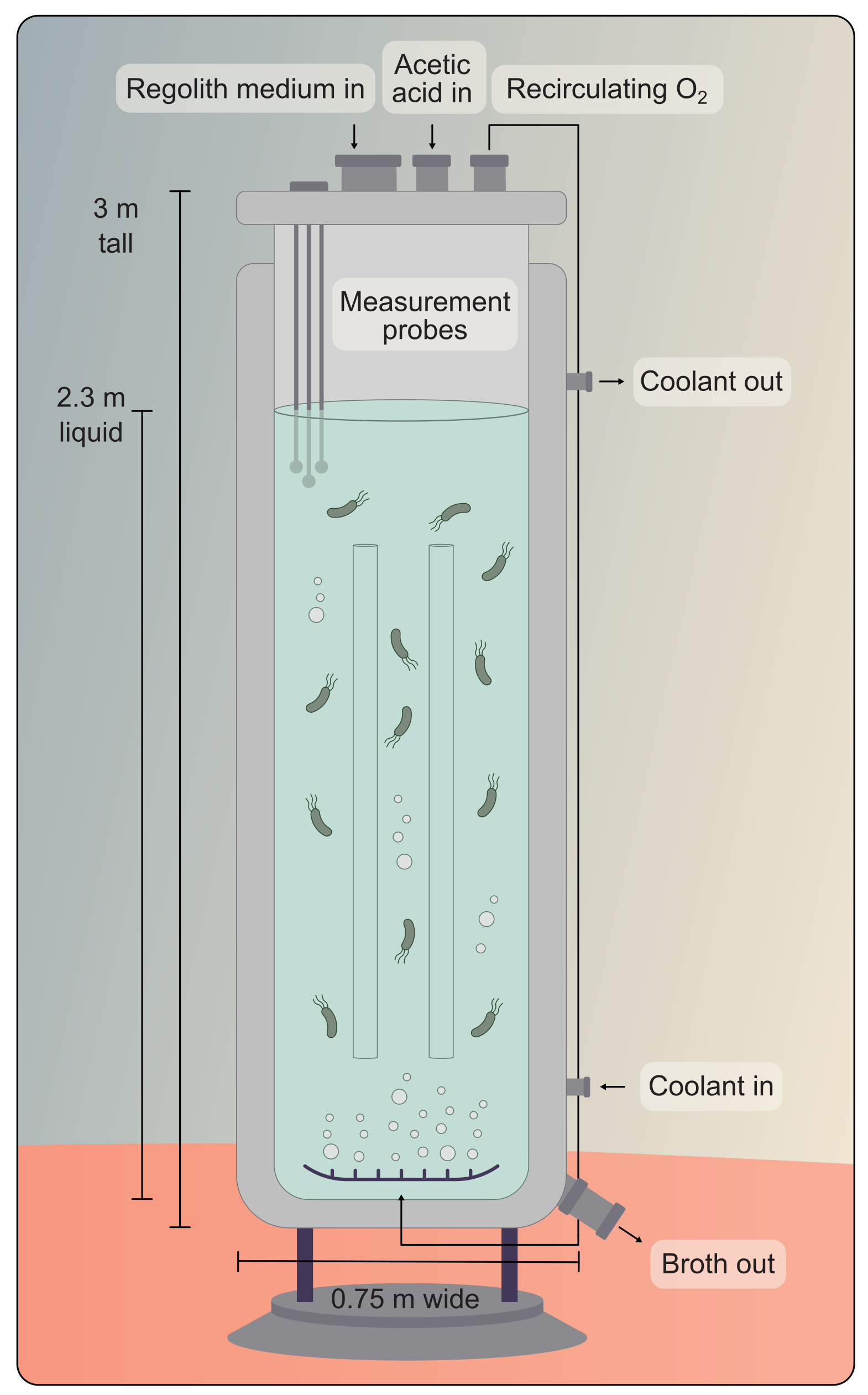}{0.36}{Bioreactor Design.}{Artistic representation of a proposed Mars surface bioreactor. Each of the four bioreactors is filled to a 1,000 L working volume with medium and inoculated with a microbial seed culture. Detection probes measure fermentation metrics, such as dissolved oxygen, pH, and optical density. Recirculated oxygen is used to maintain cell suspension. A cooling jacket enables temperature control. After the cell growth has finished, culture is removed from the bottom of the tank and dewatered for downstream processing.}
\setcounter{piofig}{4}%
\piofigureh[b]{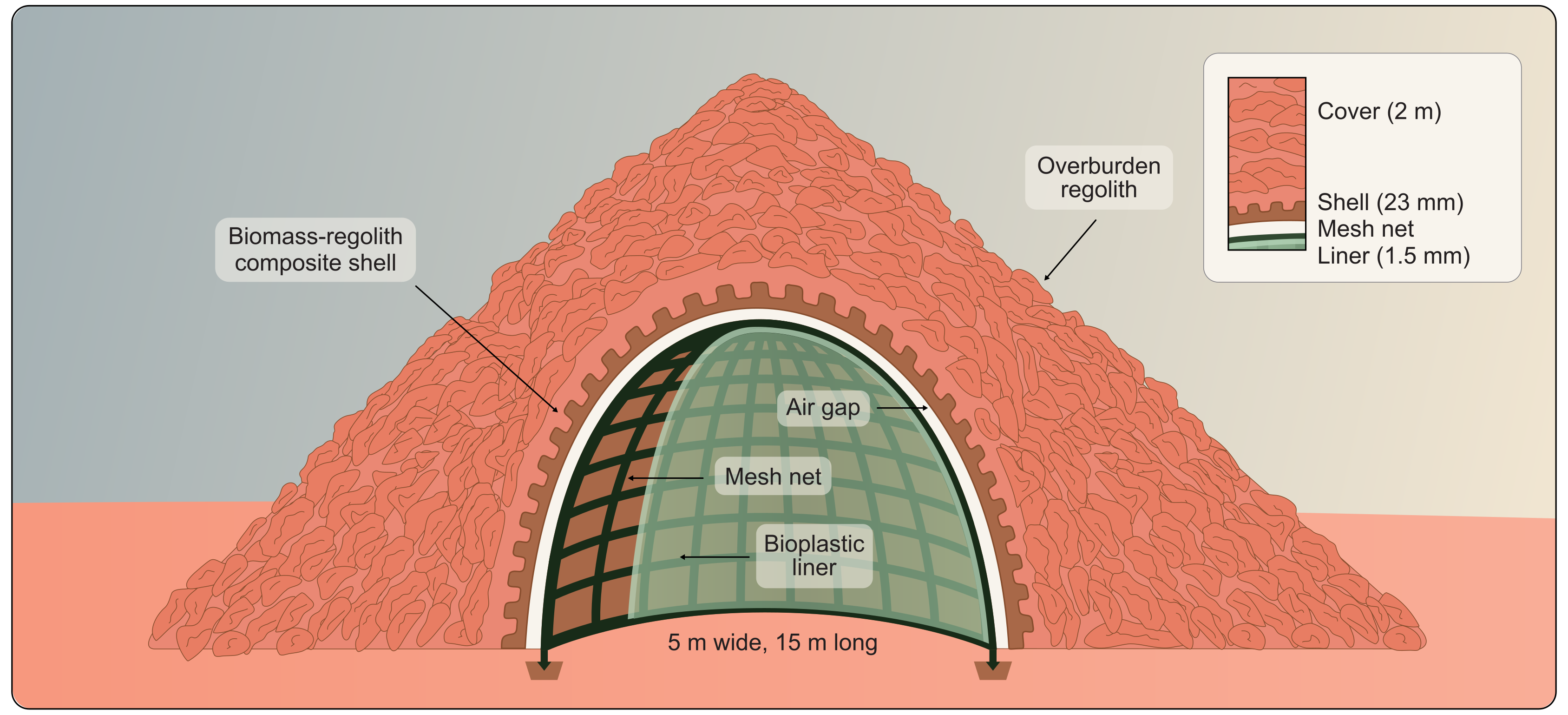}{0.22}{Mars Seed Surface Habitats.}{Artistic representation of the cross-section of regolith-biopolymer surface habitats. The habitat is constructed of two biobased components: a regolith-biomass composite exterior shell and a bioplastic interior liner. They are buried under two meters of regolith to provide radiation shielding. At 5 meters wide and 15 meters long, each habitat has sufficient livable floor space (75 m\textsuperscript{2}) to house two crew members.}
For consistency and continuous operation, the four bioreactors run in staggered sequence and are completely emptied and refilled after each cycle. Each bioreactor would run on a two-day cycle, with 29 hours to cultivate the bioplastic-rich biomass and 19 hours to empty and start a fresh fermentation process. Prior to starting a new run, the bioreactors could be opened to the Martian atmosphere to reduce the internal bioburden. It may be desirable to operate the bioreactors in a semi-continuous or continuous process, where only a portion of the culture is drained after the 29-hour growth process and refreshed with medium instead of entirely emptying the tank. This type of operation would decrease turnaround time, is common in typical biomanufacturing production processes, and has been demonstrated to increase productivity in PHA production\textsuperscript{61}. Yet it may lead to unintentional evolution that could result in inconsistent yields\textsuperscript{62,63}. For now, the safest option for modeling Mars Seed is to fully drain the tank at the end of each bioreactor’s growth process, ensuring consistent bioplastic production each cycle.

\subsubsection{Downstream Processing for Bioplastic Purification}
After the fermentation process is complete, the resulting biomass must be separated from the aqueous environment, typically referred to as dewatering. The daily output of the bioreactors is 2,000 L/day of an aqueous solution, with roughly 7 g/L of biomass at 70\% bioplastic by mass. Dewatering here is accomplished with continuous-flow centrifugation techniques that are widely used in the biomanufacturing, food, and chemical industries\textsuperscript{64}, and typically operate at large scales. For example, a small-sized commercial model processes 4,000 L/day, draws 1.5 kW of power in continuous operation, and contributes less than 200 kg of mass, with modifications to the skid\textsuperscript{65,66}. This type of system sufficiently dewaters the bioreactor outputs and produces about 14 kg/day of biomass. A flight-ready continuous dewatering system could be built without significant development for low hundreds of kilograms and one kW of continuous power.

In typical bioprocesses, further downstream processing, like purification or washing, is required to turn the harvested biomass into usable material. Once the biomass in Mars Seed is harvested, it is split into two equal streams. The first half does not require further processing and is sent as is directly to be blended with regolith and extruded into the regolith-bearing outer shell. The other stream must be further processed into pure bioplastic to make the pressure-tight liner. Since the bioplastic is trapped inside the cells, the purification requires a two-step process where cells are lysed and then the bioplastic is separated from cell debris. One option to accomplish this is to use supercritical CO\textsubscript{2}. Complete cell lysis can be accomplished by mixing biomass with liquid CO\textsubscript{2} at supercritical temperatures and then releasing pressure, which causes the CO\textsubscript{2} to infiltrate and lyse cells, sometimes aided by additional co-solvents or wash steps\textsuperscript{67}. This process liberates PHB granules from within cells, which are further separated from cell debris as unwanted lipids are washed away with the supercritical CO\textsubscript{2}, which has been demonstrated at pilot-scale\textsuperscript{68}. This takes about 1.7 kg liquid CO\textsubscript{2}/kg raw biomass, requiring 12 kg/day of supercritical CO\textsubscript{2}, or 0.04 kW of continuous power, based on achievable pump designs to extract liquid CO2 from the atmosphere at 0.08 kWh/kg\textsuperscript{69}. If higher bioplastic purity is needed, several low-consumable approaches could be investigated, such as air flotation\textsuperscript{70}, additional continuous centrifugation methods with liquid CO\textsubscript{2}\textsuperscript{67}, or air classification\textsuperscript{71}. Mars Seed is designed with the multi-stage process of lysis and separation process allotted with a few hundred kilograms of equipment and one kW of continuous power. Completely consumable-free purification of PHB is not as developed, though operating this process on Mars grants the ability to advantageously use the present conditions (i.e. CO\textsubscript{2} for an \textit{in situ} solvent, low-pressure and low-temperature atmosphere for freeze-drying), which would decrease the mass and power requirements over what would be required on Earth.

\subsection{Habitat Building}
Surface habitats to provide astronauts with a pressurized and radiation-shielded area will be required for any sustained presence on Mars. A variety of habitat designs have been proposed for this effort, ranging from shipped inflatable habitats\textsuperscript{72} to habitats built \textit{in situ}, such as regolith-based structures\textsuperscript{73} or biobased mycotectural buildings\textsuperscript{74}. The product of Mars Seed is a habitable structure that builds on those preexisting designs by utilizing \textit{in situ} resources as the core building material while relying on a feasible biological process to produce a regolith binder and inflatable liner. The habitat is designed as a half-cylinder five meters across and 15 meters long, buried under two meters of raw regolith to provide micrometeorite, thermal, and radiation shielding (\textbf{Figure 5}). The regolith-bearing layer is a corrugated shell composed of a biomass-regolith composite. Inside the structure is a pressure-containing liner made of purified bioplastic, capable of holding the NASA exploration atmosphere of about 57 kPa\textsuperscript{75} with a safety factor of four. Combining the corrugated shell and interior liner, the total mass of one habitat is about 1,300 kg of PHB, or about 1,800 kg of bioplastic-containing biomass. A 15-meter-long structure can provide 75 m² of floor space. Standards for habitable area requirements for extraterrestrial structures vary depending on the location (i.e. surface vs. outer space) and mission design\textsuperscript{76}. We therefore choose to use NASA’s Technical Standards, such as NASA-STD-3000\textsuperscript{77}, to define that one habitat is enough for two people at about 35 m\textsuperscript{2} per person. These combined bio- and \textit{in situ-}based habitats provide a livable environment for future surface missions.

\subsubsection{Constructing the Outer Shell\textbf{ }}
The exterior of the habitat is two meters of loose regolith piled on top of a structural outer shell. This outer shell is a composite made by mixing regolith with harvested biomass at an 87:13 ratio. The biomass acts as a binder for the particulate regolith; this composite is printed as a thin corrugated wall to support the loose regolith overburden. A similar material and structure has been demonstrated in the Lunar Infrastructure Asset (LINA) setting, where a mix of simulated regolith and polylactic acid (PLA) polymer was robotically extruded through a nozzle at 180°C under near-vacuum conditions\textsuperscript{78}. Comparable mixtures of 12-20\% binder in regolith have been shown to have high flexural strengths of 8 to 11 MPa\textsuperscript{79}. Furthermore, the binder can be whole-cell biomass rather than a purified polymer, as unpurified \textit{C. necator} biomass containing 70\% PHB can be directly extruded\textsuperscript{80} and can match or exceed neat PHB in flexural strength\textsuperscript{81}. The dewatered biomass, the most simple unprocessed product of Mars Seed, can directly act as a material for habitat construction without any further modifications.

The primary purpose of the habitat shell is to provide radiation shielding and therefore must be strong enough to hold the piled-on regolith overburden, which is the functionally protective material. We use 300 g/cm\textsuperscript{2} of regolith as the minimum overburden amount based on demonstrations of the radiation shielding ability of regolith\textsuperscript{82,83}. At 1,500 kg/m\textsuperscript{3} for loose regolith\textsuperscript{84,85}, two meters of overburden provides 300 g/cm\textsuperscript{2} of shielding, meaning about 400 tonnes of loose regolith would be required for each 5 x 15 m structure. Under Mars’ reduced gravity, this places a load of

\begin{center}1,500 kg/m\textsuperscript{3} x 3.7 m/s\textsuperscript{2} x 2 m = 11 kPa\end{center}

on the crown of the structure. Based on published mechanical measurements for printed regolith-polymer composites\textsuperscript{78} and a screening calculation based on classical equivalent-stiffness methods for corrugated sheets\textsuperscript{86,87}, we developed a concept-level shell consisting of a 5 mm printed wall with corrugations approximately 90 mm deep. The calculation suggests that corrugation would provide the stiffness needed to support the regolith overburden with a safety factor of 1.5, which is based on NASA’s Technical Standards on specifications for a bonded composite structure under exterior pressure\textsuperscript{88}. This calculation idealizes the wall as a periodic sheet and uses classical closed-cylinder buckling theory as a screening analogue\textsuperscript{89} but does not model additional constraints, like the actual corrugation profile, end walls, or regolith-structure interaction. We treat these dimensions as a preliminary mass-sizing basis rather than a validated structural design, and they will require further analysis and representative load testing. The shell is built as a curved surface of a half-cylinder habitat with a surface area of 120 m\textsuperscript{2}, and with the proposed corrugation has an effective thickness of 23 mm, so a total of 2.8 m\textsuperscript{3} of blended polymer and regolith material is required. Assuming an approximate composite density of 2,400 kg/m\textsuperscript{3}, based on the combined proportional densities of the mineral regolith\textsuperscript{84,85} and the bioplastic-containing biomass\textsuperscript{90,91}, the composite shell has a mass of 6,600 kg. Since the shell is only 13\% biomass, this requires 860 kg of biomass per structure mixed with 5,800 kg of regolith. A corrugated design enables resource-efficient construction of a composite shell capable of holding up the necessary amount of regolith to provide radiation shielding.

The equipment necessary to build the habitat shell would consist of an extrusion line that blends coarse and fine regolith, mixes it with whole-cell biomass, heats it, and extrudes it through a gantry-mounted nozzle. The LINA framework again provides precedent; it used a 180 kg demonstration gantry\textsuperscript{78} to print a regolith-loaded thermoplastic into the shape of a structure within a \textasciitilde{}1 m\textsuperscript{3} footprint. The gantry to print the Mars Seed habitat shells would likely be about ten-fold heavier and need to cover a space approximately a hundred times larger. While the gantry is physically large, heating the 50 kg/day of blended material to 180 °C requires approximately 0.1 kW, while an additional 1.9 kW is available for mixing and pumping the slurry and moving the extrusion nozzle. Habitat-scale construction equipment does not require any biology-specific modifications and instead will be designed based on ongoing space habitat construction technology development.

\subsubsection{Manufacturing the Interior Bioplastic Liner}
Surface habitats must hold in an artificial atmosphere to make them livable for a crew. Inside the outer shell, we propose a flexible bioplastic liner capable of containing the pressure of a breathable atmosphere inside the half-cylinder habitat. This liner is structurally separate from the outer shell, and does not rely on the weight of the regolith to maintain its structure.

The liner can be formed from two primary components, both produced from purified bioplastics. The first component is the bulk membrane, approximately 1.5 mm thick and made of simple cast bioplastic film. It can be formed as multiple redundant 0.5 mm-thick layers, each sufficiently gas-impermeable to hold the atmosphere. With an area density of 1.9 kg/m\textsuperscript{2} calculated from a density of 1,300 kg/m\textsuperscript{3}, and covering a surface area of 230 m\textsuperscript{2}, the bulk membrane requires approximately 440 kg of PHB\textsuperscript{91}. The second component of the liner are straps, which stretch across the inside of the membrane and help contain the habitat's internal pressure. These can be made of woven bioplastic filaments formed by melt-spinning\textsuperscript{92}, which has been demonstrated with PHA polymers to form fibers with 740 MPa of tensile strength\textsuperscript{93}. This is sufficient to contain an internal pressure of the 57 kPa atmosphere\textsuperscript{75}, as the stress for a 2.5 m radius habitat would be

\begin{center}57 kPa x 2.5 m = 140 kN/m\end{center}

With a required safety factor of four for human-occupied structures\textsuperscript{72}, the required strength is about 560 kN/m. One way this could be met is with a series of PHA fiber straps, spread out over the interior of the bulk membrane. To cover the necessary strength, the straps would need a total areal thickness of about 0.76 mm as

\begin{center}560 kN/m ÷ 740 MPa = 0.76 mm\end{center}

\setcounter{piofig}{5}%
\piofigureh[b]{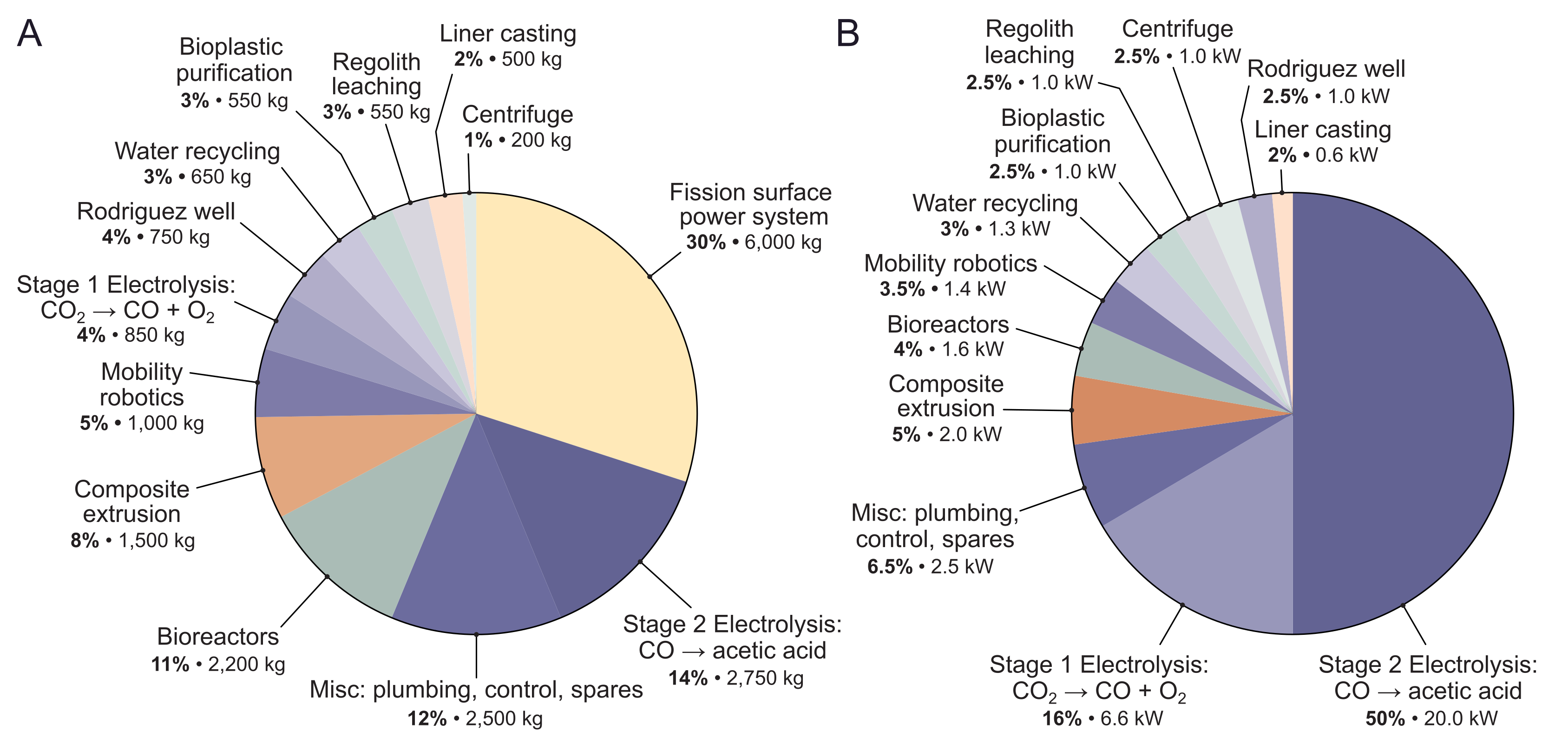}{0.3}{Mass and Power Breakdown of Mars Seed.}{\textbf{A.}~Mass of Mars Seed by subsystem. The fission surface power system accounts for the most mass (30\%), followed by acetate production. Biotic infrastructure accounts for 15\% of the mass of Mars Seed.\par\vspace{2pt}\textbf{B.}~Power draw of Mars Seed by subsystem. Two-thirds of the 40 kWe is consumed by the electrolysis subsystems for O\textsubscript{2} and acetate production. Biotic infrastructure accounts for less than 10\% of power draw.}
This effect can be achieved using a mesh net over the bulk membrane, with 1.5 mm thick straps that are 50 mm long and spaced every 100 mm, leaving 50 mm-wide gaps between the force-containing straps. They would have a density of 0.95 kg/m\textsuperscript{2} and, over a surface area of 230 m\textsuperscript{2}, would have a total mass of approximately 220 kg. The ends of these straps are anchored into the regolith to support the half-cylinder shape of the structure. The total structure of the interior two-component liner is therefore about 660 kg, with one-third in the straps and two-thirds in the bulk membrane. The two-component interior liner can be created entirely from the \textit{in situ} derivable bioplastic and is designed to hold in the breathable atmosphere at an appropriate safety factor.

The interior liner described here is a theoretical design to conceptualize feasibility, and would require further hardware development to take in 5 kg/day purified PHB, cast the bulk membrane, and melt-draw the strap fibers. The general concepts of inflatable habitats\textsuperscript{72} have been demonstrated by the BEAM module on the ISS as well as Sierra Space’s LIFE habitat, which specifically demonstrates the paired membrane and strap approach. Multiple companies, including CJ Biomaterials, Danimer Scientific, Newlight Technologies, and RWDC Industries, are improving bioproduced plastic production technology. Such hardware used on Mars will be less heavy and power-intensive relative to their Earth counterparts given the relatively small scale of the process, and could be accomplished for a few hundred kg and less than one kW of power. Significant further research and development are required before these processes could be run autonomously on Mars. Further avenues of research, such as PLA/PHA blends\textsuperscript{94} or modified monomers\textsuperscript{95} are ongoing, highlighting future work to genetically engineer production organisms to produce biopolymers with improved mechanical properties. Finally, the membrane would require rigid fittings for an airlock, valves, and structural interfaces, which are not included in this model. Such fittings would be required for any early habitat constructed on the Martian surface.

\section{Manifest and Technical Development Priorities}
Mars Seed outlines a complete end-to-end bioprocess accomplished using 13 subsystems totaling 20,000 kg, designed to maximize the bio-ISRU process while bound by the 40 kWe fission surface power system. The biofoundry subsystems have substantial overlap with other space mission architectures, allowing us to draw assumptions from the existing arsenal of design reference studies. We chose assumptions for the mass, power, throughput of each subsystem to represent a physically plausible, internally consistent system based on the best available design studies, prototypes, and scaling relationships (\textbf{Table 1}). Every subsystem still requires substantial further development before it could be used in a Mars mission context, ranging from laboratory testing of pressure-tight bioplastic liners to scale-up of MOXIE.

\setcounter{piofig}{6}%
\piofigureh[t]{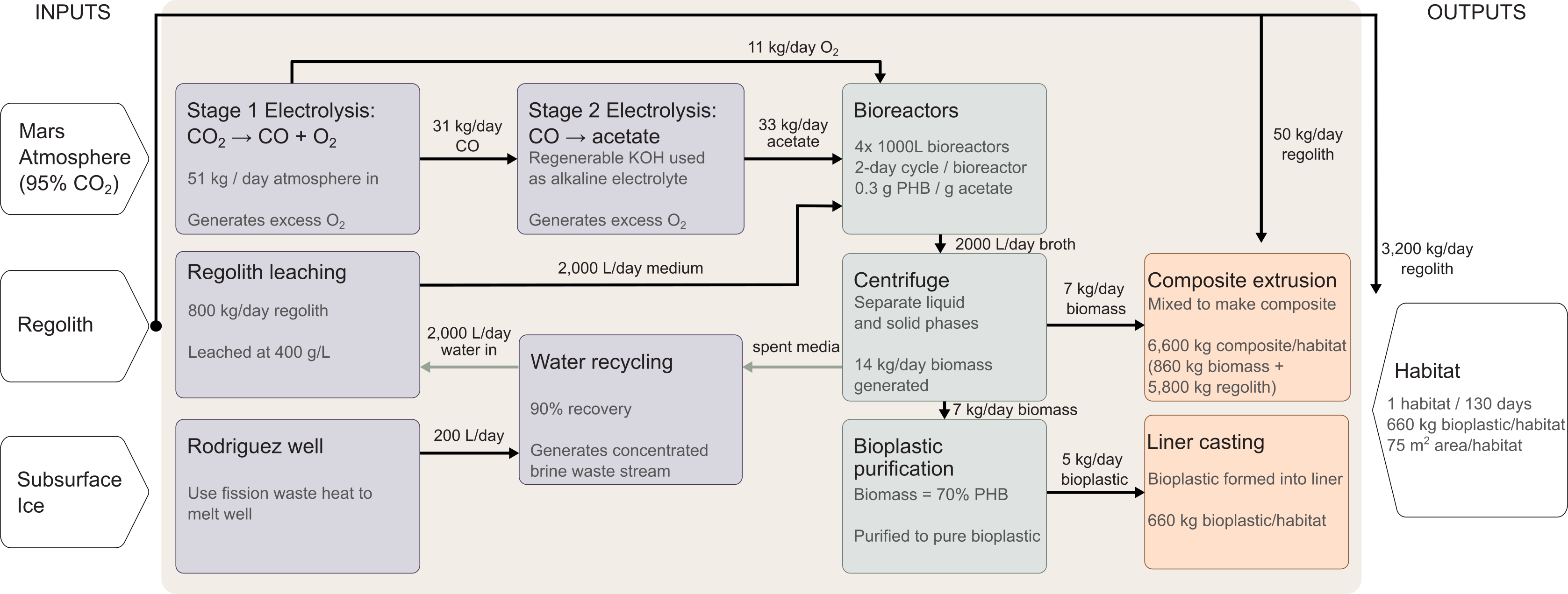}{0.26}{Mars Seed Process Flow Diagram.}{Process flow diagram detailing the daily input and outputs for each resource-transforming subsystem.}
Maturation of these technologies is already in progress. Power and water are currently high priority areas of research for space agencies, and surface fission reactors and Rodwells have been designed. CO-to-acetate electrolysis is already commercially scaled on Earth\textsuperscript{42} and would therefore need focused development to adapt it for space exploration. Heavy-lift landers and autonomous robotic operation are shared requirements of nearly every proposed human-to-Mars surface mission. Much of the biotic infrastructure equipment, like bioreactors and centrifuges, exists today at the appropriate scale for Mars Seed. However, they require translating ground equipment to flight equipment, including building with lighter materials and evaluating robustness and functionality on Mars’ surface. As with any connected system, the entire biofoundry needs to be tested for continuous autonomous operation; from equipment setup and regolith collection to bioplastic materials generation and habitat assembly.

\begin{table*}[p]
\centering
\piotablecaptionlab[\textwidth]{1}{Mars Seed Manifest.}{The complete manifest of equipment to be transported to Mars, with their allotted mass and power draw, informed by engineering precedents. Technical development priorities, highlighting critical areas of research, are provided for each subsystem}
\piotablebody
\begin{tabularx}{\textwidth}{L{0.115\textwidth}L{0.105\textwidth}L{0.042\textwidth}L{0.046\textwidth}XX}
\piohead{\textbf{Subsystem }} & \piohead{\textbf{Function}} & \piohead{\textbf{Mass (kg)}} & \piohead{\textbf{Power (kW)}} & \piohead{\textbf{Engineering Precedent}\textbf{ }} & \piohead{\textbf{Technical Development Priorities}} \\
Fission surface power system & Continuous electricity & 6,000 & 40 out & Based on NASA’s 40-kWe surface-power design objective and associated mass targets\textsuperscript{22,27}\textbf{.} & Continued development on dependable large-scale surface fission reactors and engineering to extend lifetime. \\
\hline
Stage 1 Electrolysis, CO₂ to CO + O₂ & Carbon monoxide and the aeration oxygen & 850 & 6.6 & Flight demonstrated on Mars at 12 g O\textsubscript{2}/hour\textsuperscript{5} and scaled to a terrestrial demonstration of 500 g O\textsubscript{2}/hour\textsuperscript{96}. Systems engineering analysis of scaleup is available\textsuperscript{38}\textbf{.} & While high TRL technology exists for oxygen ISRU, it would need adjustments to the larger throughput scale required. \\
\hline
Stage 2 Electrolysis, CO to acetic acid & Reduce CO to the carbon feedstock & 2,750 & 20 & Electrochemical conversion has been demonstrated at 0.7 kW scale\textsuperscript{39} and 10 kW scale\textsuperscript{97}, with larger commercial systems under development\textsuperscript{42}. Downstream acidification scaled from terrestrial bipolar-membrane electrodialysis\textsuperscript{43}. & The technology for this process needs to be transitioned from Earth- to space travel-readiness, including the ability to be robotically set up and serviced. The stack lifetime needs to be validated for long-term operation on Mars. Further development could remove the dependency on consumables (i.e. the bipolar membrane). \\
\hline
Rodriguez well (Rodwell) & Water from subsurface ice & 750 & 1.0 & Supported by NASA Rodwell preliminary assessment\textsuperscript{26}, Mars-condition melt-pool experiments and system modeling\textsuperscript{28}, integrated Mars-pressure prototype development\textsuperscript{98}, and decades of Antarctic Rodwell field heritage\textsuperscript{30}. & The Rodwell and equipment would need to be designed for appropriate scale and location, like parameterization of injected water temperature flow and water withdrawal rates. Mars poses a higher risk of boiling and water loss due to the low-pressure atmosphere. \\
\hline
Mobility robotics & General mobility and regolith haulage & 1,000 & 1.4 & Based on a commercial semi-autonomous lunar cargo-rover design\textsuperscript{20} and NASA regolith-excavation prototypes with tonne-per-day design capabilities\textsuperscript{21}. & Regolith excavators and collectors need to be tested at scale. Many large-scale autonomous robotics designed for the Moon would translate to future Mars missions. \\
\hline
Regolith leaching & Dissolve nutrients from regolith & 550 & 1.0 & Based on past mission data on regolith chemical makeup\textsuperscript{45,99} and scaled from our prior work on bench Mars-simulant leaching and biological validation of the resulting defined medium\textsuperscript{32}. & Assessment of ion abundance in a variety of regolith sites will help inform exact quantities of regolith medium required. Equipment could be modified from terrestrial mining and milling operations. \\
\hline
Water recycling & Recover \textasciitilde{}90\% of process water & 650 & 1.3 & Based on terrestrially well-established systems\textsuperscript{34,35} and by the long-duration water-recovery operation aboard the ISS\textsuperscript{33}. & Expansion of water recovery systems, like those on the ISS, to the required scale to process 2,000 L of media waste per day. \\
\hline
Bioreactors & Ferment acetate to PHB & 2,200 & 1.6 & Scaled from 5-L acetate-fed\textit{ C. necator} PHB fermentation\textsuperscript{58} and Mars-medium compatibility in the Pioneer study\textsuperscript{47}. Similar designs as terrestrial systems\textsuperscript{59,60}. & Much of this infrastructure already exists in semi-autonomous use for on-Earth biomanufacturing. These equipment require modification for flight-readiness as well as fully autonomous duty cycles, sterilization, and maintenance. \\
\hline
Centrifuge & Harvest and dewater the cell culture & 200 & 1.0 & Commercial centrifuges in the required 50-200 L/h range are common in terrestrial industry\textsuperscript{65,66}. & Requires demonstration at scale at low mass equipment and properly accounting for fluid dynamics in the low-pressure and low-gravity environment on Mars. Ideally continuous dewatering operation rather than batch operation to minimize risk of cake buildup. \\
\hline
Bioplastic purification & Purify the bioplastic polymer out of cells & 550 & 1.0 & Based on (sub-)gram-scale supercritical CO\textsubscript{2} recovery from mixed-culture biomass\textsuperscript{67,68,100}\textbf{.} & Purifying PHB out of cells and into liner-grade purity using only \textit{in situ} derived solvents is a major technology/hardware challenge that will need to be addressed to unlock the bioplastic pressure liners. It will also be driven by the required final purity of the PHB for forming into the liner components. \\
\hline
Liner casting & Form the pressure-holding liner & 500 & 0.6 & Standard plastic processing systems, like extruding, spinning, and casting equipment\textsuperscript{92,101} are also used for bioplastic\textsuperscript{102} and biocomposite applications, such as films, fibers, and sheets. & Desired material properties will dictate the necessary plastic processing equipment parameters to transform the polymer into the final liner \\
\hline
Composite extrusion & Form the biomass-regolith composite shell & 1,500 & 2.0 & Scaled from regolith–thermoplastic printing under vacuum and cryogenic conditions\textsuperscript{78}, molded regolith–polymer\textsuperscript{79} and regolith-biomass\textsuperscript{103} formulation studies, and laboratory extrusion of PHB-rich whole-cell biomass\textsuperscript{81}. & Scale up to production from laboratory miniatures to full-scale regolith/biomass shells. Print and load-test a representative full sized corrugated shell under Mars relevant conditions. \\
\hline
Miscellaneous: plumbing, control, spares & Fluid handling, automation control, communication, spares & 2,500 & 2.5 & Predominantly driven by spare membranes for Stage 2 electrolysis. Further estimating communication and control systems is beyond the level of detail of this study. They will be based on overarching precedent for general operations and interconnection of surface equipment. & Mature other subsystems to better define the challenges and requirements of integration. Continue development to minimize the requirement of spares or consumables. \\
\hline
\textbf{Total} &  & \textbf{20,000} & \textbf{40.0} &  &  \\
\end{tabularx}
\end{table*}
\setcounter{piofig}{7}%
\piofigurehcol[b]{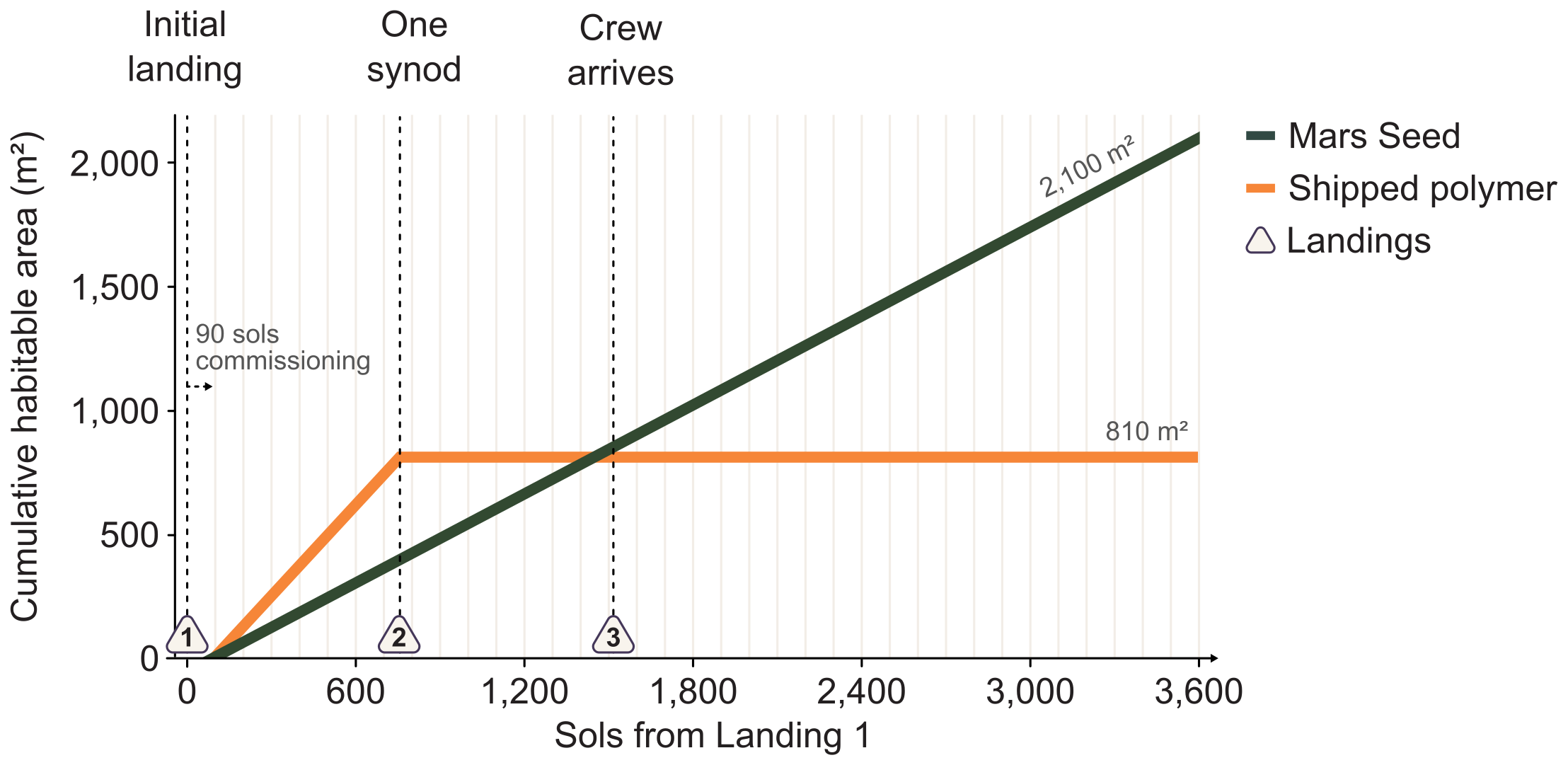}{0.26}{Mars Seed Compared to Shipping Building Material.}{Amount of habitable surface area (m\textsuperscript{2}) created by Mars Seed over a ten-year timeframe compared to an equal mass mission with shipped building material. Even if the alternative mission is designed with a faster habitat build rate, it plateaus at 810 m\textsuperscript{2} once shipped material is exhausted.}

\subsubsection{Mass and Power Breakdown}
Breaking down Mars Seed into the individual subsystems can highlight insights into massive and power-hungry functions. Looking at the mass breakdown, a third of the mass of Mars Seed is dedicated to the fission surface power system, with another third allotted for the \textit{in situ} resource gathering and conversions, and the last third comprising the biotic infrastructure, connective plumbing, and the habitat building subsystems (\textbf{Figure 6A)}. Only about 15\% of the total mass is dedicated to the biotic infrastructure, predominantly the bioreactor systems which are sized to meet the bioproduction demands, with redundancy built in. After the fission reactor, the next largest subsystem is the CO-to-acetate electrolysis system. This subsystem is also the largest energy user of the equipment in Mars Seed (\textbf{Figure 6B}) and it is one of the few that requires spare parts to be sent (bipolar membrane for KOH regeneration). The biotic equipment uses a smaller fraction of power, less than 10\%. Even as a biological ISRU mission, the majority of the mass and power in Mars Seed goes towards abiotic infrastructure and functions, much of which requires development for any surface mission with a crew presence.

\subsubsection{Production Capability}
In total, three raw \textit{in situ} resources, atmosphere, ice, and regolith, are converted into habitable structures by the complete Mars Seed process (\textbf{Figure 7}), generating one complete habitat about every 130 Earth days. About 50 kg of atmospheric CO\textsubscript{2} is captured daily from the atmosphere to be converted to the oxygen (11 kg/day) and acetate (33 kg/day) necessary to feed the microbe. After an initial large water draw to start the first process, 200 L of water is extracted from subsurface ice by the Rodwell daily, and is supplemented by an additional 1,800 L of recycled water to create the liquid media for the bioprocess. Over 4,000 kg of regolith is used every day, for generating the inorganic nutrients for the bioprocess (\textasciitilde{}800 kg/day), mixing with bioplastic-containing cells to form the corrugated shell (\textasciitilde{}50 kg/day), and finally providing the habitat with 2 meters of regolith for radiation shielding (\textasciitilde{}3,200 kg/day). Ultimately, the resulting biomass is purified and used to generate a habitat of 75 m\textsuperscript{2} livable area, large enough for about two people. All three raw inputs are effectively inexhaustible at these rates. Therefore, equipment lifetime, not resource availability, determines Mars Seed’s lifetime.

\setcounter{piofig}{8}%
\setlength{\piofigtopgap}{0pt}%
\piofigureh[b]{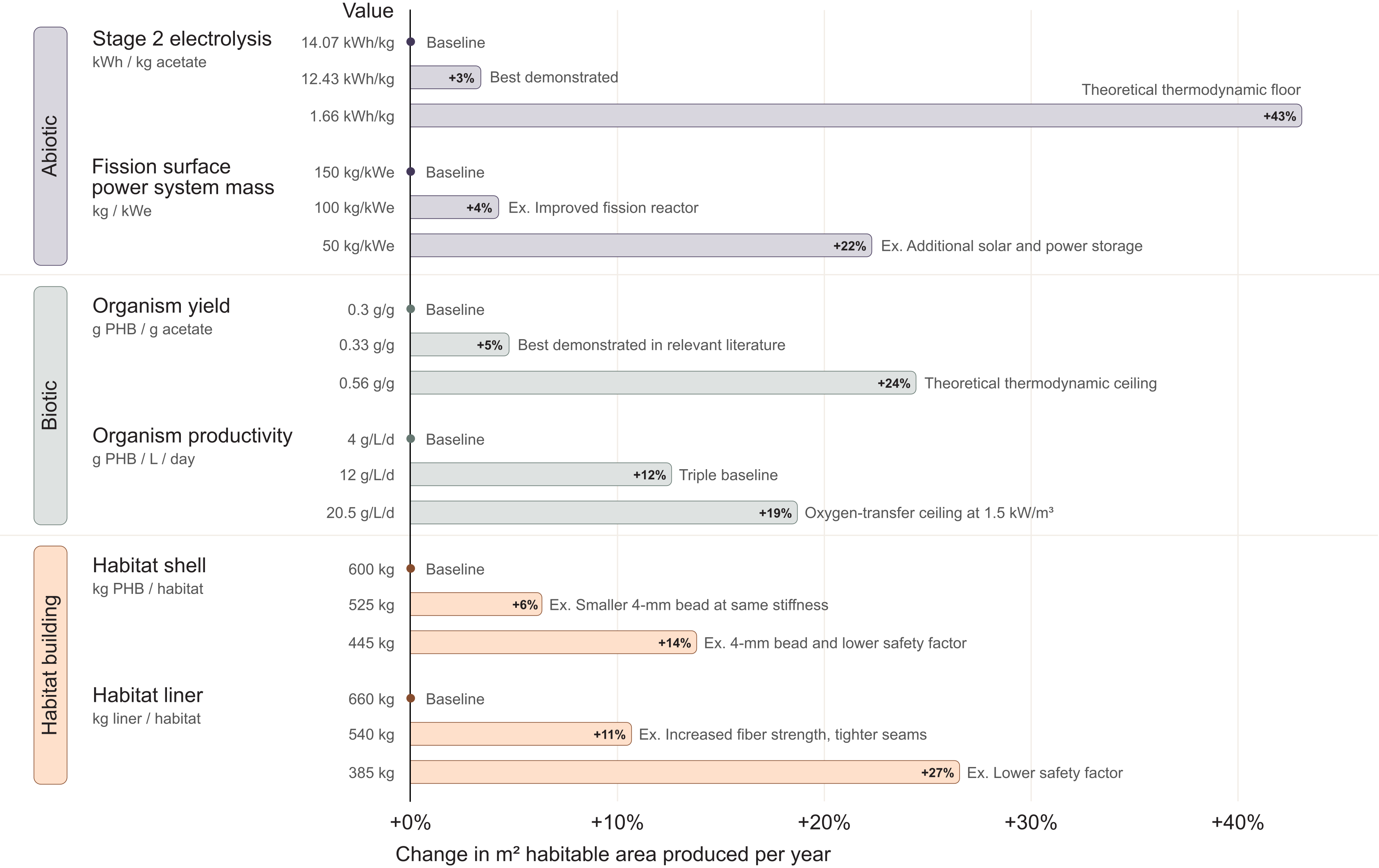}{0.44}{Sensitivity Analysis.}{A sensitivity analysis of Mars Seed highlights gains in productivity given improvements to various portions of the process. The units are given in m\textsuperscript{2} of habitable area created per year by a 20 tonne set of equipment. The first bar in each category is considered feasible given literature review, while the second bar is the maximum theoretical improvement.}
\setlength{\piofigtopgap}{12pt}%

\subsubsection{Comparison to Alternative Mission Using Shipped Construction Material}
An alternative to building the infrastructure necessary to locally produce structures is to bring in materials from Earth. Because of the mass expansion of combining a polymer with regolith to create a structure, it is significantly more mass-efficient to import a polymer, mix it with local regolith and extrude a structure than it is to bring pre-manufactured construction materials like steel\textsuperscript{104,105}. Therefore, the best comparison point for this architecture is an alternative version where a 1,000-kg FSP system powers the necessary robotics, regolith mixing and extrusion infrastructure, with a total mass of 6,500 kg (\textbf{Table S1}). Remaining mass up to 20,000 kg is then available to launch raw polymer as the building material to mix with regolith and extrude shells similar to the Mars Seed habitats, as well as pre-cast inflatable liner habitats from Earth. In this alternative mission (“Shipped polymer”), 13.5 tonnes of polymer produces a maximum of 10 complete habitats, or 810 m\textsuperscript{2} of total possible habitable area (\textbf{Figure 8}). Mars Seed produces the same yield in 1,400 sols, about 100 sols before the two-synodic period between the initial landing and the arrival of a long-term habitation crew, as in the NASEM 30-Cargo-300 campaigns. At two synodic periods, it has produced 870 m\textsuperscript{2}. Then it continues, producing 2,100 m\textsuperscript{2} of habitable area over a ten-year timeframe, the equivalent to shipping 43,000 kg of extrusion infrastructure and polymer materials to Mars. Even if the alternative mission is able to construct habitable area more quickly, once the stock of material is exhausted, it is not able to continue producing structures. Therefore, given the same mass allotment, Mars Seed is a more intriguing design that builds the same amount of habitats while derisking key science questions and providing a path towards a sustained and scalable surface presence.

\subsubsection{Sensitivity analysis}
We conducted a sensitivity analysis of Mars Seed to understand which improvements to the process would yield the greatest increases in productivity. For each of six key variables, we consider a reasonable optimization based on existing literature and also the largest justifiable improvement. We compare the productivity increase in terms of habitable area produced compared to the base Mars Seed design (220 m\textsuperscript{2} per year) (\textbf{Figure 9}). Note that accurately comparing possible optimizations requires re-optimizing the remaining subsystem requirements based on a 20 tonne fixed total mass and 40 kWe maximum power allotment for the entire mission.

The most feasible optimization of the process is improvements to the bioproduction organism, which would yield a meaningful increase in the amount of habitable area produced. For example, optimizing biological productivity would produce PHB at a faster rate, increasing the amount of yearly habitable area. Increasing the PHB yield to what has been demonstrated in literature with acetate as a feedstock\textsuperscript{58}, or the theoretical maximum conversion of acetate to PHB, would decrease the demand for more media. Potentially large improvements in the power and mass efficiencies of the fission reactor and electrolysis systems would also increase throughput, since those two subsystems are nearly half of the mass of the biofoundry. A thermodynamically maximum efficiency of the acetate-generating electrolysis would give massive gains, however the best demonstrated efficiency\textsuperscript{39} improves throughput by only a few percent. Therefore, it is unlikely that such large gains will be easily realized. Finally, decreasing the material demands for habitable structures increases the amount of habitable area produced. For example, printing the habitat shells at a smaller bead size (ex. 4 mm as demonstrated in LINA\textsuperscript{78}) or improving the strength of the bioplastic liner would decrease the required amount of bioproduct to make each habitat. Further lowering the safety factors also decreases the material demands, but is not a viable option prior to extensive stress testing of the habitat. A future combined sensitivity analysis, where multiple aspects of the process are improved or changed in parallel, could yield larger benefits, as each process's throughput could increase in step instead of shifting resources to accommodate other less-efficient processes. Overall, research focused on biological optimization provides a clear path towards improved efficiency while technological development on abiotic ISRU and extraterrestrial planet habitat construction continues.

\section{Compatibility with Other Mission Proposals}
Mars Seed is described herein as a standalone bioprocess. However, the infrastructure for Mars Seed has substantial overlap with many other missions designed for Mars (\textbf{Table 2)}. Any human mission will require robotics, power, oxygen, atmospheric compression, and water for life support and propellant generation. Many long-term missions will rely on uncrewed precursor missions to autonomously setup a surface presence. Systems engineering approaches can identify synergies between shared components and available waste streams, such as excess oxygen, brine, and biomass created by the biofoundry. The overlap in mission needs raises the potential for larger joint missions with economies of scale, integration of waste streams, and development of core technologies while accomplishing shared mission objectives.

\begin{table*}[b]
\centering
\piotablecaptionlab[\textwidth]{2}{Process and Mission Overlaps with Other Studies.}{Overlap in processes and mission objectives between Mars Seed and select similar studies. Process overlaps detail shared equipment (ex. FSP) and requirements (ex. Tonne-scale lander, autonomous infrastructure assembly). Mission overlaps detail shared techniques (ex. ISRU O\textsubscript{2} generation), timeframes (ex. Launch window timing), and objectives (ex. Surface habitat, bio-ISRU).}
\piotablebody
\begin{tabularx}{\textwidth}{L{0.20\textwidth}L{0.21\textwidth}X}
\piohead{Study} & \piohead{Process Overlaps} & \piohead{Mission Overlaps} \\
NASEM 30-Cargo-300 Campaigns (2026)\textsuperscript{4} & \textbullet~Autonomous infrastructure setup prior to crew arrival \newline \textbullet~Rodriguez well for water ISRU & Utilizing strategically designed launch windows, the 30-Cargo-300 campaign specifically could align with the two synodic period timeline of Mars Seed. Mars Seed similarly envisions a dedicated infrastructure setup mission to initiate ISRU prior to extended surface presence. Specifically, ice and regolith are highlighted as used for both basic needs (i.e. water) and as a building material (i.e. structures of bound regolith and/or ice). \\
\hline
Human Exploration of Mars Design Reference Architecture 5 (DRA 5) (2010, 2014)\textsuperscript{19,106} & \textbullet~Autonomous infrastructure setup prior to crew arrival \newline \textbullet~Fission surface power \newline \textbullet~SOXE for O\textsubscript{2} generation & Beyond using the same system for O\textsubscript{2} generation as DRA 5 (MOXIE), Mars Seed uniquely takes advantage of the carbon monoxide generated from SOXE, where an additional electrolysis step converts the typically vented CO into acetate, a biocompatible feedstock. \\
\hline
Moon to Mars (2025)\textsuperscript{13} & \textbullet~Fission surface power system \newline \textbullet~Landing of large payloads & Mars Seed details specific equipment capable of the required atmosphere and ice abiotic ISRU highlighted in Moon to Mars. Several cited desired Mars Functions or Mars Use Cases are described in Mars Seed, such as “equipment to accommodate biological experiments” (FN-U-204M) and “demonstration of construction using surface-born resources” (UC-I-201M). \\
\hline
Starship Mission Architecture  (2022)\textsuperscript{107} & \textbullet~Autonomous infrastructure setup prior to crew arrival \newline \textbullet~Fission surface power system \newline \textbullet~Rodriguez well for water ISRU & Mars Seed relies on successful completion of a Mars lander capable of a many tonne payload. Beyond demonstrating this lander, the Starship surface campaign also highlights the importance of other necessary ISRU functions for self-sufficiency (O\textsubscript{2}, water). \\
\hline
Mars Surface Habitat Concept Design (2023)\textsuperscript{108} & \textbullet~Fission surface power system \newline \textbullet~Autonomous infrastructure setup prior to first EVA \newline \textbullet~Landing of large payloads & The Mars Surface Habitat (MSH), based on a lunar habitat, is made of inflatable softgoods, much like the Mars Seed habitat liner, yet would also provide crew habitation in space. MSH calls for venting the crew’s CO\textsubscript{2} waste which could instead be pumped into a smaller setup for conversion for bio-ISRU. \\
\hline
Rocket Propellant via bio-ISRU Process Design (2021)\textsuperscript{15} & \textbullet~Basic bioprocess equipment (ex. bioreactor, cell separation) \newline \textbullet~Water recycling & Utilizing cyanobacteria as the system for O\textsubscript{2} generation would remove the need for abiotic ISRU. In that case, the nutrient rich cyanobacteria biomass can feed the Mars Seed bioprocess as an energy source in lieu of CO\textsubscript{2} electrolysis. \\
\hline
Biomanufacturing-driven Reference Mission Architecture (2022)\textsuperscript{109} & \textbullet~Atmosphere conversion for oxygen and other nutrients \newline \textbullet~Regolith collection for nutrients \newline \textbullet~Low-mass bioreactor system & Acetate is a commonly cited ISRU-derivable carbon and energy source for microbial bioprocesses. The specific bioprocess of Mars Seed (PHA generation from \textit{C. necator}) is highlighted as a possible product. All similarly depend on other \textit{in situ} resources, such as water and regolith, and standard bioprocess equipment, like bioreactors. \\
\end{tabularx}
\end{table*}
Beyond the universal objective of exploring Mars, many of the objectives of previously designed missions have core overlaps with Mars Seed. Dating back to Mars Direct\textsuperscript{69}, most designs propose utilizing uncrewed landings to assess the area and establish available resources, for both ISRU and science research. Designs for surface habitats range from near-term inflatable structures\textsuperscript{72,108} to long-term harnessing of local resources\textsuperscript{4,13}, much like Mars Seed, and have been designed specifically for Mars\textsuperscript{110} or as a general habitat design\textsuperscript{111}. Most Mars mission studies accomplish similar objectives.

The NASEM 30-Cargo-300 campaigns are especially intriguing for compatibility with Mars Seed. The 30-Cargo-300 campaigns incorporate a launch dedicated to the autonomous setup of a surface infrastructure with a sustained human presence in subsequent launches. Assuming each launch is timed with planetary alignment to minimize the travel time, the 300 portion of the campaign lands two synodic periods after the 30 launch. That time aligns with Mars Seed’s structure building timeline, where sufficient structures to house a larger crew would be established by the time of the 300 landing. As a three-launch design, the 30-Cargo-300 is uniquely suited to establish a biological ISRU process in the earliest mission, providing it time to operate prior to a longer surface presence landing.

\begin{piobox}[b]{Mini Mars Seed (MMS)}
\setlength{\columnsep}{12pt}%
\begin{wrapfigure}{r}{0.60\linewidth}
\vspace{-0.8\baselineskip}
\setcounter{piofig}{9}%
\piofigureinline{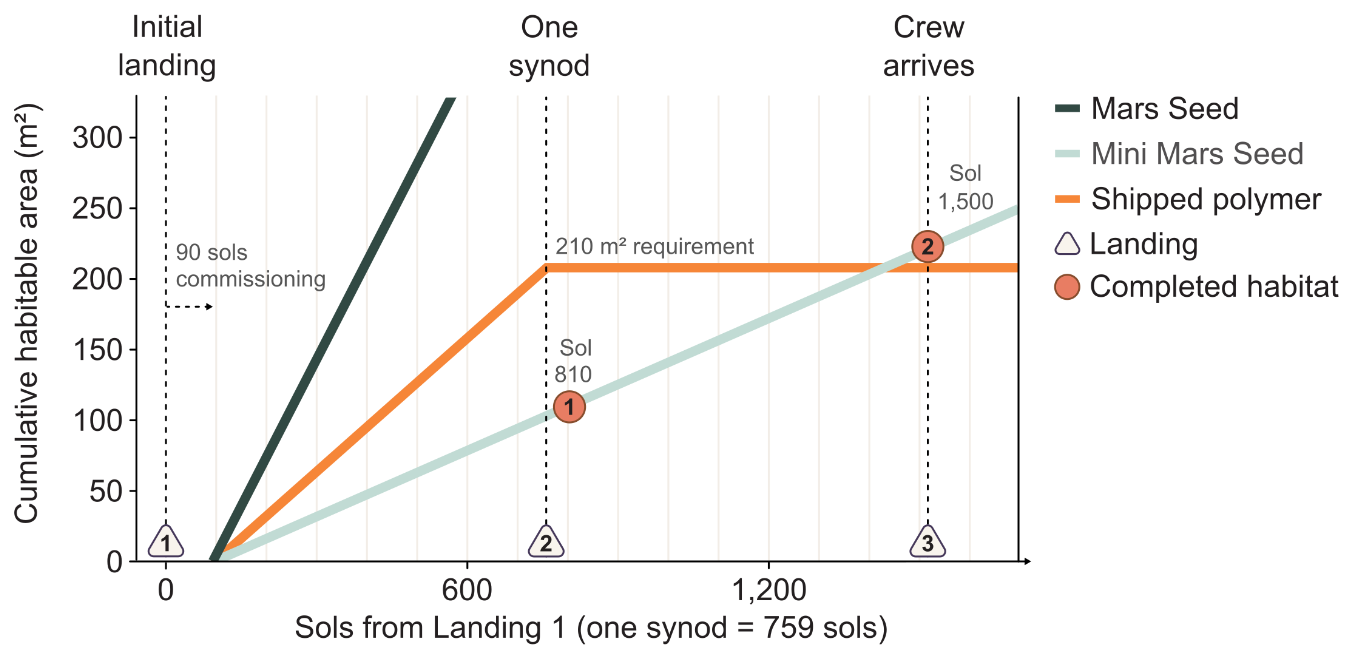}{0.40}{Mini Mars Seed Compared to Shipping Building Material.}{Amount of habitable surface area (m\textsuperscript{2}) created by Mini Mars Seed (MMS) over a five-year timeframe compared to an equal mass mission with shipped building material. Even if shipped bioplastic facilitates a faster habitat build rate, as shown, it plateaus once shipped material is exhausted.}
\end{wrapfigure}
The architecture described throughout the main text optimizes bioplastic production given a dedicated 40 kWe fission reactor, resulting in 870 m\textsuperscript{2} of habitat over two synodic periods, considerably more than is required for the proposed first six-person human mission. Alternatively, we can develop a design that meets the minimum habitation needs of an early human outpost while minimizing power and mass. Using the same standard of 35 m\textsuperscript{2} habitable area per person\textsuperscript{77}, a six-person crew, as described in the NASEM campaigns\textsuperscript{4}, requires at least 210 m\textsuperscript{2} of floor space. Assuming the 30-Cargo-300 launches each mission in optimal launch windows, the 210 m\textsuperscript{2} of livable habitat needs to be established in two synodic periods. We call this alternative sizing Mini Mars Seed (MMS), and it may be a more attractive option to include in early mission proposals.

MMS requires all of the same equipment and processes as the full design but at lower throughput and therefore lower mass and power requirements. To further fit the minimal requirements, the size of the habitats is adjusted to be 105 m\textsuperscript{2} livable area (5 m x 21 m), fitting three astronauts per habitat instead of two but resulting in slightly more bioplastic needed per habitat. Now at 1,500 kg of bioplastic per habitat, 3,000 kg of bioplastic is required in total. At the same titer of 5 g/L bioplastic, 900 L of culture generates 4.5 kg per cycle and therefore makes 3,000 kg in about 1,300 days, or 1.7 synodic periods. The culture amount drives the \textit{in situ} resource requirements and the amount of necessary daily nutrients (regolith medium, acetate, and oxygen) is decreased to only feed a single bioreactor. Similarly, if wastewater recycling is maintained at the same rate as in Mars Seed (90\%), the Rodwell only needs to supply 45 L per day, requiring much less power and a smaller mass pump. Scaling down the power and mass by the new required nutrient amounts results in MMS using around 10 kWe, 25\% of a 40 kWe fission reactor, and a total mission mass of 8,100 kg (\textbf{Table S2}). We pro-rate the fission reactor's mass contribution by power draw, allotting MMS one quarter of the reactor's 6,000 kg mass. In practice, a single full reactor would more likely be shipped to serve multiple missions rather than a fraction built for MMS alone. At 40\% of the mass of Mars Seed, MMS accomplishes the minimum amount of habitat construction using scaled-down versions of the same infrastructure.

We can compare MMS to a similarly sized alternative mission that brings raw polymer material instead of the bioprocess equipment. MMS is sized precisely to generate 210 m\textsuperscript{2} of habitable area within two synodic periods. The alternative mission could build the same required area faster as it would be limited by the habitat-building equipment rather than the biology productivity rate but would again be limited in expansion without polymer resupply from Earth (\textbf{Figure 10}). At the designed mass and power allotment, MMS becomes more attractive to include as an expansion of an existing mission that includes a larger fission reactor. While not achieving the same production goals with the larger Mars Seed architecture, improved processes may render MMS more competitive against shipping building material, or more viable for long-term scaling of a base designed for multiple missions.
\end{piobox}
Beyond mission objectives and potential launch timelines, there are several overlapping technologies between these proposals. NASA’s surface power fission concept is considered a necessary power source for surface missions\textsuperscript{22}. The development of autonomous systems and robotics is another throughline, most prominently defined in Moon to Mars architecture, and will be required for most future missions focused on establishing a sustained presence\textsuperscript{13}. The proposals involving surface presence highlight developed \textit{in situ} oxygen generation systems, such as MOXIE, which, regardless of mission, would need to be appropriately sized to meet expected oxygen demands. In addition, the NASEM 2026 Science Strategy references Rodwells and bio-chemical-physical closed-loop systems as enabling technologies\textsuperscript{4}. Lastly, most designs utilize large capacity landers, ranging from a 25 tonne requirement\textsuperscript{13,112} to Starship’s 100 tonne design\textsuperscript{107}.

A systems approach can identify ways for integrating waste streams to improve the efficiency of a complex mission architecture. For example, Mars Seed uses the CO waste stream generated by MOXIE as an input for acetate production. The biofoundry produces two potential useful waste streams. A concentrated brine solution from the biofoundry’s water recycling will be high in ions like chloride, sodium, and potassium. This concentrate could be processed by bipolar-membrane electrodialysis to produce acid and base streams, as demonstrated for terrestrial desalination brines\textsuperscript{113}. These reagents could support pH control, equipment cleaning, mineral processing, and alkali-activated regolith binders\textsuperscript{114}. The biofoundry also produces cell biomass waste from the bioplastic purification process for the liner (about 2 kg per day), which is a rich nutrient source high in biologically available nitrogen and carbon and could serve as an optimal input for more specialized bioprocesses or fertilizer for soil-based crops. Finally, the regolith leaching process produces a substantial amount of spent regolith. This washed regolith, now depleted of both the inorganic nutrients and soluble toxins (i.e. perchlorate), can be used as a building material, either as described in Mars Seed or in a similar ISRU habitat design. It remains critical to identify overlap to find opportunistic chances for making compatible mission designs that accomplish respective science objectives while using closed-loop systems and shared infrastructure.

\section{Extending into other bioprocesses}
Mars Seed draws on a biomanufacturing paradigm that is already mature at industrial scale on Earth, and that can in principle be used to produce most of the materials and products of Earth's biosphere. On Earth, microbial fermentation already produces more than 100 billion liters of ethanol\textsuperscript{115} and more than 2 million tonnes of citric acid annually\textsuperscript{116}. Microbial fermentation has been our source for recombinant human insulin since 1982\textsuperscript{117}. Terrestrial bioreactors can range from small bench-scale systems to 1,000 L pilot systems to 500,000–1,000,000 L reactors for bulk commodity fermentation\textsuperscript{63}. The space-biotechnology community has proposed using biomanufacturing in space for food, medicine, materials, and waste recycling\textsuperscript{9,10,118}. Rather than build a new industry from scratch, the goal is to adapt the infrastructure, scale, and reliability already established on Earth to deliver a large fraction of the products that will be needed in space. Mars Seed is a minimum viable version of that larger vision. It provides a complete template for end-to-end bioproduction, including both feedstock production and processing, that can be expanded to additional products.

\begin{table*}[!t]
\centering
\piotablecaptionlab[\textwidth]{3}{Open Scientific Questions.}{Select key open scientific questions relevant to Mars Seed and brief proposals of next steps and potential impacts.}
\piotablebody
\begin{tabularx}{\textwidth}{L{11pt}L{0.33\textwidth}X}
\piohead{} & \piohead{\textbf{Scientific}\textbf{ question}} & \piohead{\textbf{Potential impact and next steps}} \\
\multirow{4}{*}[-53pt]{\rotatebox[origin=c]{90}{\accentfont\fontsize{6.5}{9}\selectfont Biology}} & How does regolith composition and site-to-site variability affect fermentation biomass generation and bioplastic yields? & Site-to-site variability in regolith composition could positively or negatively impact biomass yield by altering the leached nutrients and potential stressors from the regolith\textsuperscript{132}. It is crucial to gain a better understanding of regolith composition at candidate landing sites and test fermentation performance with analog or returned samples\textsuperscript{133}. Prioritizing aqueous geochemistry experiments on Mars, as was done on the Phoenix lander\textsuperscript{45}, will provide more insight into regolith variability. \\
\cline{2-3}
 & Can the engineered organism maintain productivity and physiological stability over years of repeated reseeding and growth? & We need long-duration growth studies across repeated cycles of seeding new cultures and growth to measure productivity and organismal stability over time\textsuperscript{134}. Minimizing radiation exposure during the scale-up process should also help maintain consistency. \\
\cline{2-3}
 & Do the fermentation yield and kinetics hold at Mars Seed production scale and duty cycle? & Larger scale Mars regolith analog based fermentation needs to be tested and optimized to confirm target fermentation yields and rates. Medium simulants already exist\textsuperscript{32}, as do appropriately sized bioreactors. Relatedly, studies are needed to understand the effect of fed-batch vs. (semi)continuous operation on productivity, much like the traditional scale-up process\textsuperscript{134}. The ability to control the automated fermentation process with remote monitoring must also be considered. \\
\cline{2-3}
 & What level of cleaning is required to prevent buildup of insoluble inorganic material inside the bioreactors over years of operation? & Continuous operation of bioreactors will depend on the ability to autonomously empty and clean the inside of the tank. Testing with leached regolith simulants and different cleaning and sterilization methods will inform the rate of potential buildups. Preventative procedures and intermittent cleaning to maintain continuous operation of bioreactors is standard in biomanufacturing. \\
\hline
\multirow{5}{*}[-40pt]{\rotatebox[origin=c]{90}{\accentfont\fontsize{6.5}{9}\selectfont Engineering}} & What level of robotics is necessary to assemble and operate the biofoundry? & The number and type of robotics will be determined by the variety and frequency of tasks. For example, regolith collecting will be a continuous operation while the fission reactor is only set up once. Complete systems engineering analysis and further comparison to current or future Mars exploration proposals will indicate the minimum level of robotics. \\
\cline{2-3}
 & What is the feasibility of other production techniques for building materials on Mars? & Comparison of the feasibility and structural properties of regolith-biopolymer composite habitats to other potential \textit{in situ} regolith binders, like ice\textsuperscript{73} or sulfur\textsuperscript{135}, must be done. \\
\cline{2-3}
 & How much perchlorate remains inside the liner post-purification, and is an additional purification/remediation step required to reduce it to human-safe levels? & Presence of detectable perchlorate in the biomass or bioplastic recovered will influence the safety of the habitats. The amount of perchlorate in recovered material should be quantified to test if it meets human-safe levels. Countermeasures to perchlorate will continue to be studied\textsuperscript{136}. \\
\cline{2-3}
 & What are the gas-permeability, tensile strength, and long-term durability properties of the purified bioplastic liner? & The materials properties of the bioplastic liner need to be tested under longer-term thermal and pressure cycling. Such testing of materials is already a common part of habitable softgoods development\textsuperscript{72}. \\
\cline{2-3}
 & What are the mechanical properties of the regolith-biomass composite used in the outer shell, including UV degradation? & Understanding how extrusion or printing parameters influence material performance can be informed from 3D-printing practices\textsuperscript{137}. More work needs to be done to characterize the strength and lifetime of the load-bearing regolith-biomass composite outer shell. \\
\end{tabularx}
\end{table*}
Many additional products and organisms can be added to the Mars Seed architecture without redesigning the core processes. Bioplastics are a broad family of materials\textsuperscript{119} used in applications ranging from packaging products to textile fibers. Many bioplastics can also be chemically modified, blended, or copolymerized to expand their material properties\textsuperscript{120}. With genetic engineering, \textit{C. necator} can produce a number of different bioplastics, such as longer chain length PHAs or PHA co-polymers, like with PLA, that expand the potential physical properties of the resulting material to include flexible or rubber-like polymers\textsuperscript{121–123}. \textit{C. necator} can also produce commodity chemicals like isopropanol and the terpene $\beta$-farnesene\textsuperscript{124–126}. The same acetate-fed bioreactor could also be used to grow other acetate-utilizing microorganisms\textsuperscript{41,127,128} with only modest changes to the bioreactor’s operational conditions, such as \textit{Corynebacterium }\textit{glutamicum} to produce recombinant proteins\textsuperscript{129} and \textit{Escherichia coli} to produce 2,3-butanediol and acetoin\textsuperscript{130}. Across all of these, the reactor, feedstock, and parts of the downstream process (ex. cell lysis and separation) built for Mars Seed's PHB production carries over unchanged.

Larger modifications would open up more product options, at the expense of added complexity or reliance on finite consumables. First, replacing acetate with glucose is a direct route to higher yields, since sugar is more readily used by cells for growth. However, Mars has no established \textit{in situ} source of sugar, and producing it locally would require more abiotic infrastructure than Mars Seed’s CO₂-to-acetate system. Second, hydrogen-oxidizing microbes could consume H₂ and CO₂ directly to produce food and biopolymers\textsuperscript{131}, as have been previously proposed for space biomanufacturing\textsuperscript{11}. This removes the need for acetate and its associated infrastructure, but introduces other equipment complexity, including requiring generating, storing, delivering, and recycling H₂. Third, photosynthesis could be used to fix carbon, eliminating electrochemical fixation that dominates Mars Seed’s mass and power budget. However, it creates a different engineering problem: that of supplying and maintaining a large illuminated and heated growing area\textsuperscript{14,15,40}. Finally, relaxing the requirement for consumable-free purification would greatly expand the range of products that could be recovered at high yield and application-grade purity, especially recombinant proteins and pharmaceuticals. However, purification consumables (resins, membranes, solvents, adsorbents) can account for 35–77\% of process inputs in modeled pharmaceutical purification bioprocesses in space\textsuperscript{18} thus requiring either shipping consumables from Earth or building the infrastructure to manufacture or regenerate it on Mars. Throughout, Mars Seed serves as the starting point: a complete worked example, from feedstock to a purified product, built on a quantifiable and comparatively small footprint of equipment. Additional bioprocesses can be built on top of the analysis done here and compared to Mars Seed using similar metrics of mass/power allotment and biological ISRU productivity.

\section{Open Scientific Questions}
There are many open scientific questions that could greatly impact this architecture’s mass and power targets, as well as its realization (\textbf{Table 3}). Current research efforts have focused on derisking core scientific principles of biological ISRU, with the required characteristics of biobased habitats as only one of many research areas. Some concepts and equipment can be studied in a Moon-to-Mars setting, such as autonomous robotics and regolith harvest, while others need to be investigated directly on Mars for ISRU technologies, such as oxygen and water generation. The biomanufacturing concepts, such as bioreactor design and cell separation equipment, are well established in on-Earth industrial scale fermentation and need to be tested in the space exploration context. New discoveries or advances in biomanufacturing, chemical processes, and materials science will be necessary to build out the technology for Mars Seed.

\section{Conclusion}
Here, we present Mars Seed, an end-to-end reference architecture that uses a bioprocess to convert Martian atmosphere, ice, and regolith into pressure-containing, radiation-shielded habitats. A 20-tonne biofoundry operating with a self-contained 40-kWe power budget produces 870 m² of habitable space within approximately two synodic periods. Once deployed, the process requires no recurring Earth-supplied consumables and can continue producing habitats as long as the equipment remains functional.

Our analysis shows that the bioprocess cannot be evaluated in isolation. The biotic infrastructure equipment represents only about 15\% of the landed mass and less than 10\% of the power demand. Most of the mass and power is dedicated to abiotic infrastructure that supports the organism: notably fission power used to fix atmospheric carbon. The system-level costs of biological production become clear only when the mass and power requirements of all accompanying infrastructure are explicitly accounted for. Doing so allows biological ISRU to be directly compared to alternatives using common measures of mass, power, time, and output.

Mars Seed is envisioned as a starting framework to guide future work. It makes explicit the assumptions that underlie a functional consumables-free bioprocess for Mars, enabling scrutiny and optimization of key technical bottlenecks. Key uncertainties include acetate-electrolysis stack lifetimes and consumable-free PHB recovery. These assumptions define concrete research and engineering priorities deserving future work. By explicitly accounting for feedstock generation and product purification, Mars Seed provides a framework for determining what other terrestrial bioprocesses must operate independent of Earth supply chains. Similarly, mission designers can use Mars Seed for understanding bio-ISRU integration opportunities that outperform other approaches. Future studies can use Mars Seed as a common baseline by replacing individual organisms, products, unit operations, or habitat designs and measuring the consequences for the complete system.

\section{Acknowledgements}
This research was funded in part by The Astera Institute. In compliance with Astera’s Open Science policy\textsuperscript{138}, this paper will not be submitted to a journal and is presented here in its final form. We thank Rachel Sevey for her assistance with scientific communication. We thank Edwin Kite, Adrian Dumitrescu, and Alex King for helpful discussions and feedback. AI assistants were used in the research and design of this paper and were used to generate early versions of figures.

\newpage

\section{Supplemental Information}
\begin{table}[!h]
\centering
\piotablecaptionlabx[\columnwidth]{S1}{Shipped Polymer Mission Manifest.}{Alterations to the manifest that could be used if shipping bioplastic polymer from Earth, rather than producing it on Mars.}
\piotablebody
\begin{tabularx}{\columnwidth}{L{0.19\columnwidth}XL{0.10\columnwidth}L{0.10\columnwidth}}
\piohead{Subsystem} & \piohead{Function} & \piohead{Mass (kg)} & \piohead{Power (kW)} \\
Mobility robotics & General mobility and regolith haulage & 1,000 & 1.4 \\
\hline
Composite extrusion & Shipped polymer combined with regolith extruded into structural habitat shells & 1,500 & 2 \\
\hline
Liner assembly & Form the pressure-holding liner & 500 & 0.6 \\
\hline
Miscellaneous: plumbing, control, spares & Automation control, communication, spares & 2,500 & 2.5 \\
\hline
Fission surface power system & Continuous electricity; prorated 6.5 kWe at 150 kg/kWe & 1,000 & – \\
\hline
Bioplastic Polymer & Building material & 13,500 & – \\
\hline
\multicolumn{2}{l}{Total landed} & 20,000 & 6.5 \\
\end{tabularx}
\end{table}
\begin{table}[!h]
\centering
\piotablecaptionlabx[\columnwidth]{S2}{Mini Mars Seed equipment manifest.}{The complete manifest of equipment for Mini Mars Seed, with their allotted mass.}
\piotablebody
\begin{tabularx}{\columnwidth}{XL{0.22\columnwidth}}
\piohead{\textbf{Subsystem}} & \piohead{\textbf{Mass (kg) }} \\
Fission surface power system (prorated) & 1,500 \\
\hline
Stage 2 Electrolysis, CO to acetic acid & 1,500 \\
\hline
Stage 1 Electrolysis, CO₂ to CO + O₂ & 170 \\
\hline
Bioreactor & 530 \\
\hline
Centrifuge, bioplastic purification & 110 \\
\hline
Rodriguez well & 450 \\
\hline
Mobility robotics & 650 \\
\hline
Regolith leaching & 240 \\
\hline
Water recycling & 150 \\
\hline
Habitat building equipment & 1,100 \\
\hline
Misc: plumbing, control, spares & 1,700 \\
\hline
\textbf{Total} & 8,100 \\
\end{tabularx}
\end{table}
\clearpage

\section{References}

\begin{pioneerrefsnum}
\item Baker, D. \& Zubrin, R. Combining near-term technologies to achieve a two-launch manned Mars mission. in 26th Joint Propulsion Conference (American Institute of Aeronautics and Astronautics, Reston, Virginia, 1990). doi:10.2514/6.1990-1896.
\item Hoffman, S. J. \& Kaplan, D. I. Human Exploration of Mars: The Reference Mission of the NASA Mars Exploration Study Team. (1997).
\item Rucker, M. A. et al. NASA’s Strategic Analysis Cycle 2021 (SAC21) Human Mars Architecture. in 2022 IEEE Aerospace Conference (AERO) (IEEE, 2022). doi:10.1109/aero53065.2022.9843237.
\item Committee on a Science Strategy for the Human Exploration of Mars, Space Studies Board, Division on Engineering and Physical Sciences \& National Academies of Sciences, Engineering, and Medicine. A science strategy for the human exploration of mars. Preprint at 10.17226/28594 (2026).
\item Hoffman, J. A. et al. Mars Oxygen ISRU Experiment (MOXIE)-Preparing for human Mars exploration. Sci. Adv. 8, eabp8636 (2022).
\item Sanders, G. \& Kleinhenz, J. Update on NASA’s ISRU development and mission plans for the Artemis program. in 27th IAA Symposium on Human Exploration of the Solar System 24–37 (International Astronautical Federation (IAF), Paris, France, 2024).
\item National Academies of Sciences, Engineering, and Medicine; Division on Engineering and Physical Sciences; Aeronautics and Space Engineering Board; Space Studies Board; Committee on Biological and Physical Sciences Research in Space 2023-2032. Thriving in space: Ensuring the future of biological and physical sciences research: A decadal survey for 2023-2032. (2023).
\item Verseux, C., Poulet, L. \& de Vera, J.-P. Editorial: Bioregenerative life-support systems for crewed missions to the Moon and Mars. Front. Astron. Space Sci. 9, (2022).
\item Santomartino, R. et al. Toward sustainable space exploration: a roadmap for harnessing the power of microorganisms. Nat. Commun. 14, 1391 (2023).
\item Averesch, N. J. H. et al. Microbial biomanufacturing for space-exploration-what to take and when to make. Nat. Commun. 14, 2311 (2023).
\item Nangle, S. N. et al. The case for biotech on Mars. Nat. Biotechnol. 38, 401–407 (2020).
\item Wordsworth, R. et al. Applied astrobiology: An integrated approach to the future of life in space. Astrobiology 25, 327–330 (2025).
\item National Aeronautics and Space Administration. Moon to Mars Architecture Definition Document ESDMD-001 – Revision C. (2025).
\item Berliner, A. J. et al. Towards a Biomanufactory on Mars. Front. Astron. Space Sci. 8, (2021).
\item Kruyer, N. S., Realff, M. J., Sun, W., Genzale, C. L. \& Peralta-Yahya, P. Designing the bioproduction of Martian rocket propellant via a biotechnology-enabled in situ resource utilization strategy. Nat. Commun. 12, 6166 (2021).
\item Joris, P. et al. Assessing the Recycling Potential of Cupriavidus necator for Space Travel: Production of SCPs and PHAs from Organic Wastes. in (2023 International Conference on Environmental Systems, 2023).
\item Menezes, A. A., Cumbers, J., Hogan, J. A. \& Arkin, A. P. Towards synthetic biological approaches to resource utilization on space missions. J. R. Soc. Interface 12, 20140715 (2015).
\item McNulty, M. J. et al. Evaluating the cost of pharmaceutical purification for a long-duration space exploration medical foundry. Front. Microbiol. 12, 700863 (2021).
\item Drake, B. G. \& Watts, K. D. Human Exploration of Mars Design Reference Architecture 5.0 Addendum \#2. (2014).
\item FLEX Rover – Astrolab. \url{https://www.astrolab.space/flex-rover/}.
\item Mueller, R. P. et al. Design of an Excavation Robot: Regolith Advanced Surface Systems Operations Robot (RASSOR) 2.0. (2016).
\item Kaldon, L. K., Presby, A. L., Mason, L. S., Smith, B. K. \& Weaver, A. S. Current Status of NASA’s Fission Surface Power Project. (2024).
\item Bramson, A. M. et al. Widespread excess ice in Arcadia Planitia, Mars. Geophys. Res. Lett. 42, 6566–6574 (2015).
\item Stuurman, C. M. et al. SHARAD detection and characterization of subsurface water ice deposits in Utopia Planitia, Mars. Geophys. Res. Lett. 43, 9484–9491 (2016).
\item Clifford, S. M. et al. Depth of the Martian cryosphere: Revised estimates and implications for the existence and detection of subpermafrost groundwater. J. Geophys. Res. 115, (2010).
\item Hoffman, S. J., Andrews, A. \& Watts, K. “‘Mining’ Water Ice on Mars: An Assessment of ISRU Options in Support of Future Human Missions. NASA Johnson Space Center (2016).
\item Oleson, S. et al. A Deployable 40 kWe Lunar Fission Surface Power Concept. in (2022).
\item Hoffman, S. J., Lever, J. H., Andrews, A. D. \& Watts, K. D. Mars Rodwell Experiment Final Report. (2020).
\item Mining. Water Ice on Mars An Assessment of ISRU Options in Support of Future Human Missions.
\item Lunardini, V. Thermal design of an Antarctic water well. NSF Ice Drilling Program.
\item Stroble, S. T., McElhoney, K. M. \& Kounaves, S. P. Comparison of the Phoenix Mars Lander WCL soil analyses with Antarctic Dry Valley soils, Mars meteorite EETA79001 sawdust, and a Mars simulant. Icarus 225, 933–939 (2013).
\item Greene, H. et al. Defined Mars Media (DMM), a chemically defined simulant of the soluble macro- and micro- nutrients in Mars regolith for use in biological research. bioRxiv 2026.04.24.719001 (2026) doi:10.64898/2026.04.24.719001.
\item Carter, L. \& Schunk, G. ISS Water Recovery System, Vapor Compression Distillation Process in Microgravity. (2019).
\item Hernández-Baño, P., Calleja-Cayón, P., Vera-García, F. \& Molina-García, A. Key operational variables in mechanical vapor compression for Zero Liquid Discharge processes: Performance and efficiency evaluation. Sustainability 17, 9212 (2025).
\item Zhou, Y., Shi, C. \& Dong, G. Analysis of a mechanical vapor recompression wastewater distillation system. Desalination 353, 91–97 (2014).
\item Zhang, P. et al. Chem-bio interface design for rapid conversion of CO2 to bioplastics in an integrated system. Chem 8, 3363–3381 (2022).
\item Elwell, J. et al. Overcoming Technical Challenges to Advance the MOXIE Solid Oxide Electrolysis Stack from TRL 3 to 9, Early Challenges to Successful ISRU Oxygen Production on Mars. in (51st International Conference on Environmental Systems, 2022).
\item MacDonald, A. K., Olivares, J. F., Li, Y. \& Hollist, M. Integration of solid oxide co-electrolysis of steam and CO2 in future closed-loop life support systems. in 54th International Conference on Environmental Systems (International Conference on Environmental Systems, 2025). doi:10.32865/2346/102589.
\item Crandall, B. S. et al. Kilowatt-scale tandem CO2 electrolysis for enhanced acetate and ethylene production. Nat. Chem. Eng. 1, 421–429 (2024).
\item Crandall, B. S., Harland-Dunaway, M., Jinkerson, R. E. \& Jiao, F. Electro-agriculture: Revolutionizing farming for a sustainable future. Joule 8, 2974–2991 (2024).
\item Crandall, B. S., Overa, S., Shin, H. \& Jiao, F. Turning carbon dioxide into sustainable food and chemicals: How electrosynthesized acetate is paving the way for fermentation innovation. Acc. Chem. Res. 56, 1505–1516 (2023).
\item Hutchings, G. Lectrolyst Announces New Carbon Conversion (C2) Capsule. Lectrolyst Livewire \url{https://livewire.lectrolyst.com/p/lectrolyst-announces-new-carbon-conversion-c2-capsule} (2025).
\item Patil, R. A. Production of acetic acid in kraft pulp mill biorefinery using bi-polar membrane electrodialysis. (2016).
\item Zhu, P. et al. Direct and continuous generation of pure acetic acid solutions via electrocatalytic carbon monoxide reduction. Proc. Natl. Acad. Sci. U. S. A. 118, e2010868118 (2021).
\item Hecht, M. H. et al. Detection of perchlorate and the soluble chemistry of martian soil at the Phoenix lander site. Science 325, 64–67 (2009).
\item Rzymski, P., Poniedziałek, B., Hippmann, N. \& Kaczmarek, Ł. Screening the survival of Cyanobacteria under perchlorate stress. Potential implications for Mars in situ resource utilization. Astrobiology 22, 672–684 (2022).
\item Martin, F. R. et al. Bioplastic Production from simulated 100\% in situ Mars resources. bioRxiv (2026) doi:10.64898/2026.09.14.748363.
\item Hakami, M. W. et al. Ceramic microfiltration membranes in wastewater treatment: Filtration behavior, fouling and prevention. Membranes (Basel) 10, E248 (2020).
\item Wu, J. et al. Process intensification in stirred tanks. Chem. Eng. Technol. 35, 1125–1132 (2012).
\item Bosecker, K. Bioleaching: metal solubilization by microorganisms. FEMS Microbiol. Rev. 20, 591–604 (1997).
\item Baláž, P. \& Dutková, E. Fine milling in applied mechanochemistry. Miner. Eng. 22, 681–694 (2009).
\item Ming, D. \& Morris, R. Chemical, mineralogical, and physical properties of martian dust and soil. 1966, 6027 (2017).
\item Zwietering, T. N. Suspending of solid particles in liquid by agitators. Chem. Eng. Sci. 8, 244–253 (1958).
\item Nattermann, U. et al. A framework for evaluating forward contamination risk for bio-ISRU microorganisms. arXiv [astro-ph.IM] (2026).
\item Weldon, M. \& Euler, C. Physiology-informed use of Cupriavidus necator in biomanufacturing: a review of advances and challenges. Microb. Cell Fact. 24, 30 (2025).
\item Zhang, L. et al. A review on enhancing Cupriavidus necator fermentation for poly(3-hydroxybutyrate) (PHB) production from low-cost carbon sources. Front. Bioeng. Biotechnol. 10, 946085 (2022).
\item Nygaard, D., Yashchuk, O., Noseda, D. G., Araoz, B. \& Hermida, É. B. Improved fermentation strategies in a bioreactor for enhancing poly(3-hydroxybutyrate) (PHB) production by wild type Cupriavidus necator from fructose. Heliyon 7, e05979 (2021).
\item Kedia, G., Passanha, P., Dinsdale, R. M., Guwy, A. J. \& Esteves, S. R. Evaluation of feeding regimes to enhance PHA production using acetic and butyric acids by a pure culture of Cupriavidus necator. Biotechnol. Bioprocess Eng. 19, 989–995 (2014).
\item Chisti, Y. \& Moo-Young, M. Bioreactor design. Basic biotechnology 3, 181–200 (2006).
\item Spier, M. R., Vandenberghe L P D, P., M. A. B. \& Soccol, C. R. Application of different types of bioreactors in bioprocesses. Bioreactors: design, properties and applications 4, 55–90 (2011).
\item Santolin, L., Waldburger, S., Neubauer, P. \& Riedel, S. L. Substrate-flexible two-stage fed-batch cultivations for the production of the PHA copolymer P(HB-co-HHx) with Cupriavidus necator Re2058/pCB113. Front. Bioeng. Biotechnol. 9, 623890 (2021).
\item Rugbjerg, P., Myling-Petersen, N., Porse, A., Sarup-Lytzen, K. \& Sommer, M. O. A. Diverse genetic error modes constrain large-scale bio-based production. Nat. Commun. 9, 787 (2018).
\item Stanbury, P. F., Whitaker, A. \& Hall, S. J. Principles of Fermentation Technology. (Butterworth-Heinemann, Oxford, England, 2016).
\item Beveridge, T. CENTRIFUGATION | Large-Scale Centrifugation. in Encyclopedia of Separation Science 320–329 (Elsevier, 2000).
\item Lab disc centrifuge. \url{https://www.koenmachine.com/sys-pd/33.html}.
\item Alfa Laval. \url{https://www.alfalaval.us/products/separation/centrifugal-separators/separators/}.
\item Haghighi, F. H. et al. Supercritical carbon dioxide-based approach for the recovery and purification of polyhydroxyalkanoates from mixed microbial cultures: A green approach for bioplastics production. J. Supercrit. Fluids 228, 106760 (2026).
\item Yılmaz Nayır, T., Konuk, S. \& Kara, S. Extraction of polyhydroxyalkanoate from activated sludge using supercritical carbon dioxide process and biopolymer characterization. J. Biotechnol. 364, 50–57 (2023).
\item Zubrin, R., Baker, D. \& Gwynne, O. Mars direct - A simple, robust, and cost effective architecture for the Space Exploration Initiative. in 29th Aerospace Sciences Meeting (American Institute of Aeronautics and Astronautics, Reston, Virginia, 1991). doi:10.2514/6.1991-329.
\item van Hee, P., Elumbaring, A. C. M. R., van der Lans, R. G. J. M. \& Van der Wielen, L. A. M. Selective recovery of polyhydroxyalkanoate inclusion bodies from fermentation broth by dissolved-air flotation. J. Colloid Interface Sci. 297, 595–606 (2006).
\item Noda, I. Process for recovering polyhydroxyalkanotes using air classification. US Patent (1998).
\item Valle, G. D., Litteken, D. \& Jones, T. C. Review of habitable softgoods inflatable design, analysis, testing, and potential space applications. in AIAA Scitech 2019 Forum (American Institute of Aeronautics and Astronautics, Reston, Virginia, 2019). doi:10.2514/6.2019-1018.
\item Reches, Y. Concrete on Mars: Options, challenges, and solutions for binder-based construction on the Red Planet. Cem. Concr. Compos. 104, 103349 (2019).
\item Rothschild, L. J. et al. Mycotecture off planet: Fungi as A building material on the moon and mars. in Lunar and Planetary Science Conference (ntrs.nasa.gov, 2022).
\item Norcross, J. et al. Effects of the 8 psia / 32\% O 2 atmosphere on the human in the spaceflight environment. (2013).
\item Caiazzo, A., Casaburo, A., Petrone, G., De Rosa, S. \& Franco, F. Structures for extraterrestrial habitat: A review. Acta Astronaut. 238, 471–493 (2026).
\item Connolly, J. H. \& Arch, M. NASA Standard 3000 Human Systems Integration Standards (HSIS) update. Proc. Hum. Factors Ergon. Soc. Annu. Meet. 49, 2018–2022 (2005).
\item Gelino, N. J. et al. Application of regolith polymer composite fused granular fabrication construction in simulated lunar conditions. in Earth and Space 2024 844–857 (American Society of Civil Engineers, Reston, VA, 2024).
\item Chen, T. et al. Inorganic–organic hybrid of lunar soil simulant and polyethylene. J. Mater. Civ. Eng. 28, 06015013 (2016).
\item Collet, C., Vaidya, A. A., Gaugler, M., West, M. \& Lloyd-Jones, G. Extrusion of PHA-containing bacterial biomass and the fate of endotoxins: A cost-reducing platform for applications in molding, coating and 3D printing. Mater. Today Commun. 33, 104162 (2022).
\item Mitantsoa, J. T. et al. Microbial production of thermoplastic-targeted biopolymer composites made of polyhydroxybutyrate and protein by Cupriavidus necator CECT 4623 strain: Bacterial accumulation, thermal characterizations, direct extrusion processing and mechanical properties. Mater. Today Commun. 45, 112277 (2025).
\item Slaba, T. C., Mertens, C. J. \& Blattnig, S. R. Radiation Shielding Optimization on Mars. (2013).
\item Gakis, D. \& Atri, D. Modeling the effectiveness of radiation shielding materials for astronaut protection on Mars. arXiv [astro-ph.EP] (2022) doi:10.48550/arXiv.2205.13786.
\item Cannon, K. M., Britt, D. T., Smith, T. M., Fritsche, R. F. \& Batcheldor, D. Mars global simulant MGS-1: A Rocknest-based open standard for basaltic martian regolith simulants. Icarus 317, 470–478 (2019).
\item Lewis, K. W. et al. A surface gravity traverse on Mars indicates low bedrock density at Gale crater. Science 363, 535–537 (2019).
\item Rao, K. P. Shear buckling of corrugated composite panels. Compos. Struct. 8, 207–220 (1987).
\item Briassoulis, D. Equivalent orthotropic properties of corrugated sheets. Comput. Struct. 23, 129–138 (1986).
\item Structural Design and Test Factors of Safety for Spaceflight Hardware..
\item Hilburger, M. W. Buckling of thin-walled circular cylinders. (2020).
\item Bratbak, G. \& Dundas, I. Bacterial dry matter content and biomass estimations. Appl. Environ. Microbiol. 48, 755–757 (1984).
\item Laycock, B., Halley, P., Pratt, S., Werker, A. \& Lant, P. The chemomechanical properties of microbial polyhydroxyalkanoates. Prog. Polym. Sci. 39, 397–442 (2014).
\item Kopf, S., Åkesson, D. \& Skrifvars, M. Textile fiber production of biopolymers – A review of spinning techniques for polyhydroxyalkanoates in biomedical applications. Polym. Rev. (Phila. Pa) 63, 200–245 (2023).
\item Tanaka, T., Yabe, T., Teramachi, S. \& Iwata, T. Mechanical properties and enzymatic degradation of poly[(R)-3-hydroxybutyrate] fibers stretched after isothermal crystallization near Tg. Polym. Degrad. Stab. 92, 1016–1024 (2007).
\item Gallo, G., Piccoli, E., Bombardi, L., Aulitto, M. \& Fusco, S. Enzymatic and microbial routes to bioplastics: The green chemistry frontier of biopolymers. FEBS Open Bio 16, 709–726 (2026).
\item Cywar, R. M. et al. Elastomeric vitrimers from designer polyhydroxyalkanoates with recyclability and biodegradability. Sci. Adv. 9, eadi1735 (2023).
\item Hartvigsen, J., Hollist, M., Elwell, J., Elangovan, S. \& Voecks, G. Mission-scale MOXIE development driven prospects for ISRU and atmosphere revitalization. in 52nd International Conference on Environmental Systems (International Conference on Environmental Systems, 2023). doi:10.32865/2346/94712.
\item Hutchings, G. Lectrolyst Unveils Record 10 kW CO2-to-Acetate System. Lectrolyst Livewire \url{https://livewire.lectrolyst.com/p/lectrolyst-unveils-record-10-kw-co2-to-acetate-system} (2025).
\item Mellerowicz, B. et al. RedWater: Water mining system for Mars. New Space 10, 166–186 (2022).
\item Kounaves, S. P. et al. The MECA wet chemistry laboratory on the 2007 Phoenix Mars scout lander. J. Geophys. Res. 114, (2009).
\item Yu, J. \& Chen, L. X. L. Cost-effective recovery and purification of polyhydroxyalkanoates by selective dissolution of cell mass. Biotechnol. Prog. 22, 547–553 (2006).
\item Hufenus, R., Reifler, F. A., Fernández-Ronco, M. P. \& Heuberger, M. Molecular orientation in melt-spun poly(3-hydroxybutyrate) fibers: Effect of additives, drawing and stress-annealing. Eur. Polym. J. 71, 12–26 (2015).
\item Lim, L.-T., Auras, R. \& Rubino, M. Processing technologies for poly(lactic acid). Prog. Polym. Sci. 33, 820–852 (2008).
\item Mannes Guesser De Oliveira, S., Rezwan, K., Verseux, C. \& Maas, M. 3D printing of composites of Martian regolith simulants and cyanobacterial biomass towards sustainable material production on Mars. NPJ Microgravity 11, 60 (2025).
\item Malekpour, F., Abdali, M., Skonieczny, K., Azami, M. \& Hojjati, M. Sustainable lunar additive manufacturing of high regolith-loaded PEKK composites for space infrastructure. Compos. B Eng. 310, 113176 (2026).
\item Azami, M. et al. A comprehensive review of Lunar-based manufacturing and construction. arXiv [astro-ph.IM] (2024) doi:10.48550/arXiv.2408.05823.
\item Drake, B. G., Hoffman, S. J. \& Beaty, D. W. Human exploration of Mars, Design Reference Architecture 5.0. in 2010 IEEE Aerospace Conference (IEEE, 2010). doi:10.1109/aero.2010.5446736.
\item Heldmann, J. L. et al. Mission architecture using the SpaceX Starship vehicle to enable a sustained human presence on Mars. New Space 10, 259–273 (2022).
\item Hughes, R., McGraw, D. D., Kessler, P. D. \& Choate, A. J. Mars Surface Habitat Concept Design. in ASCEND 2023 (American Institute of Aeronautics and Astronautics, Reston, Virginia, 2023). doi:10.2514/6.2023-4750.
\item Berliner, A. J. Design and Optimization of a Biomanufacturing-Driven Reference Mission Architecture for the Human Exploration of Mars. (2022).
\item Cichan, T. et al. Mars Base Camp: An architecture for sending humans to Mars. New Space 5, 203–218 (2017).
\item Howard, R. L., Jr. Common Habitat base camp for Moon and Mars surface operations. in AIAA ASCEND 2021 (2021).
\item Goodliff, K. E., Troutman, P., Craig, D. A., Caram, J. \& Herrmann, N. Evolvable Mars Campaign 2016 - A Campaign Perspective. in AIAA SPACE 2016 (American Institute of Aeronautics and Astronautics, Reston, Virginia, 2016). doi:10.2514/6.2016-5456.
\item Reig, M., Casas, S., Gibert, O., Valderrama, C. \& Cortina, J. L. Integration of nanofiltration and bipolar electrodialysis for valorization of seawater desalination brines: Production of drinking and waste water treatment chemicals. Desalination 382, 13–20 (2016).
\item Alexiadis, A., Alberini, F. \& Meyer, M. E. Geopolymers from lunar and Martian soil simulants. Adv. Space Res. 59, 490–495 (2017).
\item Shaw, A. J. et al. Metabolic engineering of microbial competitive advantage for industrial fermentation processes. Science 353, 583–586 (2016).
\item Ciriminna, R., Meneguzzo, F., Delisi, R. \& Pagliaro, M. Citric acid: emerging applications of key biotechnology industrial product. Chem. Cent. J. 11, 22 (2017).
\item Office of the Commissioner. 100 Years of Insulin. U.S. Food and Drug Administration \url{https://www.fda.gov/about-fda/fda-history-exhibits/100-years-insulin} (2024).
\item Koehle, A. P., Brumwell, S. L., Seto, E. P., Lynch, A. M. \& Urbaniak, C. Microbial applications for sustainable space exploration beyond low Earth orbit. NPJ Microgravity 9, 47 (2023).
\item Lackner, M., Mukherjee, A. \& Koller, M. What are ‘bioplastics’? Defining renewability, biosynthesis, biodegradability, and biocompatibility. Polymers (Basel) 15, 4695 (2023).
\item Kumar, V., Sehgal, R. \& Gupta, R. Blends and composites of polyhydroxyalkanoates (PHAs) and their applications. Eur. Polym. J. 161, 110824 (2021).
\item Hectors, W., Delmulle, T. \& Soetaert, W. K. Metabolic engineering strategies for enhanced polyhydroxyalkanoate (PHA) production in Cupriavidus necator. Polymers (Basel) 17, 2104 (2025).
\item Morlino, M. S. et al. Cupriavidus necator as a platform for polyhydroxyalkanoate production: An overview of strains, metabolism, and modeling approaches. Biotechnol. Adv. 69, 108264 (2023).
\item Choi, S. Y. et al. Microbial polyhydroxyalkanoates and nonnatural polyesters. Adv. Mater. 32, e1907138 (2020).
\item Grousseau, E., Lu, J., Gorret, N., Guillouet, S. E. \& Sinskey, A. J. Isopropanol production with engineered Cupriavidus necator as bioproduction platform. Appl. Microbiol. Biotechnol. 98, 4277–4290 (2014).
\item Milker, S. \& Holtmann, D. First time $\beta$-farnesene production by the versatile bacterium Cupriavidus necator. Microb. Cell Fact. 20, 89 (2021).
\item Wang, Y. et al. Unlocking the potential of Cupriavidus necator H16 as a platform for bioproducts production from carbon dioxide. World J. Microbiol. Biotechnol. 40, 389 (2024).
\item Ma, X. et al. C2 feedstock-based biomanufacturing of value-added chemicals. Curr. Opin. Biotechnol. 73, 240–245 (2022).
\item Kiefer, D., Merkel, M., Lilge, L., Henkel, M. \& Hausmann, R. From acetate to bio-based products: Underexploited potential for industrial biotechnology. Trends Biotechnol. 39, 397–411 (2021).
\item Kiefer, D., Tadele, L. R., Lilge, L., Henkel, M. \& Hausmann, R. High-level recombinant protein production with Corynebacterium glutamicum using acetate as carbon source. Microb. Biotechnol. 15, 2744–2757 (2022).
\item Ricci, L. et al. Metabolic engineering of E. coli for enhanced diols production from acetate. ACS Synth. Biol. 14, 1204–1219 (2025).
\item Nangle, S. N., Sakimoto, K. K., Silver, P. A. \& Nocera, D. G. Biological-inorganic hybrid systems as a generalized platform for chemical production. Curr. Opin. Chem. Biol. 41, 107–113 (2017).
\item Fackrell, L. E., Humphrey, S., Loureiro, R., Palmer, A. G. \& Long-Fox, J. Overview and recommendations for research on plants and microbes in regolith-based agriculture. npj Sustain. Agric. 2, 1–23 (2024).
\item McSween, H. Y., Jr, Hamilton, V. E. \& Farley, K. A. Perspectives on Mars Sample Return: A critical resource for planetary science and exploration. Proc. Natl. Acad. Sci. U. S. A. 122, e2404248121 (2025).
\item Hewitt, C. J. \& Nienow, A. W. The scale-up of microbial batch and fed-batch fermentation processes. Adv. Appl. Microbiol. 62, 105–135 (2007).
\item Li, H. et al. Development of a novel material and casting method for in situ construction on Mars. Powder Technol. 390, 219–229 (2021).
\item Wang, J. L. et al. Potential health impacts, treatments, and countermeasures of martian dust on future human space exploration. GeoHealth 9, e2024GH001213 (2025).
\item Alkabbanie, R., Aktas, B., Demircan, G. \& Yalcin, S. Short carbon fiber-reinforced PLA composites: influence of 3D-printing parameters on the mechanical and structural properties. Iran. Polym. J. 33, 1065–1074 (2024).
\item Astera Open Science Policy. \url{https://zenodo.org/records/17873235}.
\end{pioneerrefsnum}

\end{document}